\documentclass[aip,jcp,12pt]{revtex4-1}

\usepackage{graphicx}
\usepackage{amsmath}
\usepackage{color}
\usepackage{enumerate}
\usepackage{multirow}
\usepackage{threeparttable}

\usepackage[printfigures]{figcaps}
\usepackage{mathrsfs}

\let\baraccent=\= 
\renewcommand{\=}[1]{\stackrel{#1}{=}} 

\newcommand{\E}[1]{\left< #1 \right>} 

\begin{document}


\title{Nonequilibrium path-ensemble averages for symmetric protocols}

\author{Trung Hai Nguyen}
\email{nguyentrunghai@tdtu.edu.vn}
\affiliation{Laboratory of Theoretical and Computational Biophysics, Ton Duc Thang University, Ho Chi Minh City, Vietnam}
\affiliation{Faculty of Applied Sciences, Ton Duc Thang University, Ho Chi Minh City, Vietnam. \url{nguyentrunghai@tdtu.edu.vn}.}
\author{Van Ngo}
\affiliation{Center for Nonlinear Studies, Los Alamos National Lab, Los Alamos, New Mexico}
\author{Jo\~{a}o Paulo Castro Zerba}
\affiliation{Department of Chemistry, Illinois Institute of Technology, Chicago, IL 60616, USA}
\affiliation{Institute of Physics ``Gleb Wataghin'', University of Campinas, SP 13083-859, Brazil}
\author{Sergei Noskov}
\affiliation{SYN: Centre for Molecular Simulation, Department of Biological Sciences, University of Calgary, Alberta, Canada.}
\author{David D. L. Minh}
\email{dminh@iit.edu}
\affiliation{Department of Chemistry, Illinois Institute of Technology, Chicago, IL 60616, USA. \url{dminh@iit.edu}.}


\begin{abstract}
According to the nonequilibrium work relations, path-ensembles generated by irreversible processes in which a system is driven out of equilibrium according to a predetermined protocol may be used to compute equilibrium free energy differences and expectation values. Estimation has previously been improved by considering data collected from the reverse process, which starts in equilibrium in the final thermodynamic state of the forward process and is driven according to the time-reversed protocol. Here, we develop a theoretically rigorous statistical estimator for nonequilibrium path-ensemble averages specialized for symmetric protocols, in which forward and reverse processes are identical. The estimator is tested with a number of model systems: 
a symmetric 1D potential, an asymmetric 1D potential, the unfolding of deca-alanine, separating a host-guest system, and translocating a potassium ion through a gramicidin A ion channel. When reconstructing free energies using data from symmetric protocols, the new estimator outperforms existing rigorous unidirectional and bidirectional estimators, converging more quickly and resulting in smaller error. However, in most cases, using the bidirectional estimator with data from a forward and reverse pair of asymmetric protocols outperforms the corresponding symmetric protocol and estimator with the same amount of simulation time. Hence, the new estimator is only recommended when the bidirectional estimator is not feasible or is expected to perform poorly. The symmetric estimator has similar performance to a unidirectional protocol of half the length and twice the number of trajectories.
\end{abstract}


\maketitle

\section{Introduction}

Remarkably, the nonequilibrium work theorems show that 
averages over path-ensembles generated by a special class of irreversible processes ---
when a system is driven out of equilibrium 
with a time-dependent external potential that varies according to a predetermined protocol ---
are exactly related to 
\emph{equilibrium} free energy differences \cite{Jarzynski1997,Jarzynski1997a} 
and expectation values \cite{Crooks2000}.
These theorems have been adapted to the interpretation of 
single-molecule pulling experiments \cite{Hummer2001,Hummer2005,Minh2006a,Minh2007a,Minh2008a,Minh2008,Minh2009,Hummer2010} and 
analogous steered molecular dynamics simulations \cite{Park2003}.
Moreover, they have been applied to calculating expectation values \cite{Minh2011b} and 
free energy differences \cite{Hummer2001c,Ytreberg2004c,Dellago2014}
in other contexts, including noncovalent binding free energies \cite{Sandberg2015, Giovannelli2017}.

Even if statistical estimators of equilibrium properties based on the nonequilibrium work relations are formally correct and asymptotically unbiased, they may be impractical to use due to convergence issues.
Convergence may be slow because each realization is exponentially weighted 
by the work done on the system during the nonequilibrium process, $W$.
If there is small variance in work, then this requirement has little influence on convergence \cite{Gore2003}.
On the other hand, if the variance in work is large, 
the number of trajectories necessary to obtain accurate estimates can become prohibitively large \cite{Gore2003};
exponential averages are dominated by rare events in which the work done on the system is particularly small compared to its average \cite{Jarzynski2006}. Indeed, the work must be less than the free energy difference between end states, $W \leq \Delta F$,
such that the entropy decreases.

Although entropy-reducing events are rare for unbiased sampling from a nonequilibrium path-ensemble, 
they can be more readily obtained via the conjugate twin, or time reversal, of a realization of the reverse process.
A process is defined by a protocol $\Lambda = \lambda(t)$ that specifies how parameters that control the Hamiltonian, $H(x, \lambda)$, vary with time, $0 \leq t \leq \tau$.
Every process has a unique counterpart known as the \emph{reverse} process,
$\tilde{\Lambda} = \lambda(\tau - t)$.
Each realization of a nonequilibrium driving process leads to a phase space trajectory $\Gamma = \left\{ q(t), p(t) \right\}$. 
For every trajectory, there is a time-reversed counterpart $\tilde{\Gamma} = \left\{q(\tau - t), -p(\tau - t)\right\}$ known as its \emph{conjugate twin}.
Jarzynski observed that the conjugate twin of a typical trajectory in the reverse process is a dominant trajectory in work-weighted exponential averages over forward path-ensembles \cite{Jarzynski2006}.

Inspired by Jarzynski's insight \cite{Jarzynski2006}, Minh and Adib (MA) developed a bidirectional estimator for nonequilibrium path-ensemble averages \cite{Minh2008}. 
It is bidirectional because it employs data from both forward and reverse processes;
conjugate twin trajectories from the reverse process are reweighted and combined with trajectories from the forward process.
The MA estimator has advantages over other bidirectional estimators \cite{Kosztin2006a, Chelli2009a, Frey2015, Ngo2016} because it is asymptotically unbiased,
makes no assumptions about the work distribution,
and may be used to estimate arbitrary nonequilibrium path-ensemble averages.
The first article on the estimator \cite{Minh2008} built on 
Hummer and Szabo's method for analyzing single-molecule pulling experiments \cite{Hummer2001,Hummer2005} 
to create a bidirectional estimator for two quantities: 
the free energy as a function of the harmonic trap position; 
and the potential of mean force as a function of a measured collective variable.
The bidirectional estimator was observed to have significantly less statistical bias than its unidirectional counterpart.
Subsequently, Minh and Chodera demonstrated the optimality of the MA estimator and showed how to compute its asymptotic variance \cite{Minh2009}. 
In another article, they generalized the method to the calculation of arbitrary thermodynamic expectations \cite{Minh2011b}.
Other authors have built upon the MA estimator \cite{Calderon2009, Hummer2010, Ngo2016} and applied it to diverse problems including 
permeation of a potassium ion through the pore of the ion channel gramicidin A \cite{Calderon2009, Giorgino2011, Ngo2016},
translation of a drug along the active site gorge of the enzyme acetylcholinesterase \cite{Sinha2012},
and a Diels-Alder reaction in water and methanol \cite{Soto-Delgado2016}.

Also inspired by Jarzynski's insight \cite{Jarzynski2006}, our present contribution considers a special class of nonequilibrium driven processes with a symmetric protocol. A symmetric protocol has the property that $\lambda(t) = \lambda(\tau - t)$. Such a protocol may arise from natural symmetry in a system or by appending the time reversal of a protocol onto the end of an asymmetric protocol.
An example of the former scenario 
is translocation of an ion through the symmetric gramicidin A pore; 
if the center of the membrane bilayer is at $z = 0$,
thermodynamic states with harmonic biases 
at $z = 3$ \AA~and $z = -3$ \AA~are equivalent.
A similar situation arises in simulating the translocation of a potential drug 
across a symmetric lipid bilayer.
The latter scenario in which a protocol is made symmetric may arise, for example, in single-molecule pulling experiments with atomic force microscopy in which motion of a cantilever that mechanically unfolds an RNA hairpin is immediately followed by reversing the motion of the cantilever and allowing the molecule to refold.

A unique and potentially advantageous property of symmetric protocols is that the forward and reverse processes are equivalent. Therefore, both a trajectory and its conjugate twin may be used to estimate the same nonequilibrium path-ensemble average. Another advantage of symmetric protocols is that because the initial and final thermodynamic state are the same, the free energy difference between end states is zero, $\Delta F = 0$. This implies that, unlike in the MA estimator, an estimate of the free energy difference between the initial and final states is not required to compute the dissipated work, $W - \Delta F$, the reweighting factor for conjugate twin trajectories (c.f. Equation \ref{eq:path_ensemble_integral}).

An outline of this paper is as follows:
first, we will describe a new estimator that exploits the aforementioned advantages of symmetric protocols;
computational methods for our demonstrative applications will be described;
we will present data comparing the symmetric estimator, MA estimator, and unidirectional estimators in a number of model systems;
results will be discussed;
and key conclusions will be summarized.

\section{Theory}

\subsection{Estimators of Path-Ensemble Averages}

The path-ensemble average of a functional $\mathcal{F}[\Gamma]$ that depends on a path $\Gamma$ is defined as,
\begin{eqnarray}
\label{path_ensemble_av}
\E{\mathcal{F}} = \frac{\int \mathcal{F}[\Gamma] \rho_F[\Gamma] d\Gamma}
{\int \rho_F[\Gamma] d\Gamma} ,
\end{eqnarray}
where $\rho_F[\Gamma]$ is the probability density of the path in the forward process.
For simplicity $\rho_F[\Gamma]$ incorporates 
both the probability of the phase space point in the initial state 
and of the propagation through time for the duration of the process.
Integrals are over all possible paths.
Each integral in Equation \ref{path_ensemble_av} can be doubled and expressed as a sum,
\begin{eqnarray}
\label{eq:path_ensemble_linear_combination}
\E{\mathcal{F}} = 
\frac{\int \mathcal{F}[\Gamma] \rho_F[\Gamma] d\Gamma + 
\int \mathcal{F}[\Gamma] \rho_F[\Gamma] d\Gamma}
{\int \rho_F[\Gamma] d\Gamma +
\int \rho_F[\Gamma] d\Gamma},
\end{eqnarray}

For one of the integrals in each of sums, the integral over the path $\Gamma$ can be substituted for by an integral over the path $\tilde{\Gamma}$.
Assuming that the integrator is symplectic and because the initial and final states are the same, 
the Jacobian for the transformation is unity. The substitution leads to,
\begin{eqnarray}
\label{eq:path_ensemble_substitution}
\E{\mathcal{F}} = 
\frac{\int \mathcal{F}[\Gamma] \rho_F[\Gamma] d\Gamma + 
\int \mathcal{F}[\tilde{\Gamma}] \rho_F[\tilde{\Gamma}] d\tilde{\Gamma}}
{\int \rho_F[\Gamma] d\Gamma +
\int \rho_F[\tilde{\Gamma}] d\tilde{\Gamma}}.
\end{eqnarray}

As a key step towards deriving his eponymous fluctuation theorem, Crooks \cite{Crooks1998,Crooks2000} showed that probability densities in forward and reverse path-ensembles are related by,
\begin{eqnarray}
\label{eq:path_rho_ratio}
\frac{\rho_F[\Gamma]}{\rho_R[\tilde{\Gamma}]} = e^{\beta \left( W - \Delta F \right)},
\end{eqnarray}
where $\beta = (k_B T)^{-1}$ is the inverse product of Boltzmann's constant $k_B$ and the temperature $T$, $W$ is the total work done on the system during the forward process, and $\Delta F$ 
is the change in free energy between the initial and final thermodynamic state of the forward process. For the symmetric protocols considered in this work, the free energy difference is zero, $\Delta F = 0$. Also, because forward and reverse processes are equivalent, $\rho_F[\Gamma] = \rho_R[\Gamma]$. Therefore, in the special case of symmetric protocols, Equation \ref{eq:path_rho_ratio} may be simplified into,
\begin{eqnarray}
\label{eq:path_rho_ratio_simplified}
\frac{\rho_F[\Gamma]}{\rho_F[\tilde{\Gamma}]} = e^{\beta W },
\end{eqnarray}
Substituting Equation \ref{eq:path_rho_ratio_simplified} into Equation \ref{eq:path_ensemble_substitution} leads to,
\begin{eqnarray}
\label{eq:path_ensemble_integral}
\E{\mathcal{F}} = 
\frac{\int \left[ \mathcal{F}[\Gamma] + \mathcal{F}[\tilde{\Gamma}] e^{-\beta W} \right] \rho_F[\Gamma] d\Gamma}
{\int \left[ 1 + e^{-\beta W} \right] \rho_F[\Gamma] d\Gamma},
\end{eqnarray}
An estimator for the path-ensemble average, $\overline{\mathcal{F}}$, 
may be obtained by multiplying and dividing by $\int \rho_F[\Gamma] d\Gamma$ and using the sample mean for expectation values in both the numerator and denominator,
\begin{eqnarray}
\label{eq:estimator_s}
\overline{\mathcal{F}} = 
\frac{\sum\limits_{n=1}^N \left[ \mathcal{F}[\Gamma_n] + \mathcal{F}[\tilde{\Gamma}_n] e^{-\beta W[\Gamma_n]} \right] }
{\sum\limits_{n=1}^N \left[ 1 + e^{-\beta W[\Gamma_n]} \right] },
\end{eqnarray}
where $\Gamma_n$ is the $n^{th}$ of $N$ sampled trajectories.

We will compare Equation \ref{eq:estimator_s} with two other statistical estimators for path-ensemble averages: the unidirectional estimator,
\begin{eqnarray}
\label{eq:estimator_u}
\overline{\mathcal{F}} = \frac{1}{N}
\sum\limits_{n=1}^{N} \mathcal{F}[\Gamma_n],
\end{eqnarray}
and the bidirectional (MA) estimator \cite{Minh2008, Minh2009},
\begin{eqnarray}
\label{eq:estimator_b}
\overline{\mathcal{F}} = 
\sum\limits_{n=1}^{N_f} \frac{\mathcal{F}[\Gamma_n]}
{N_f + N_r e^{-\beta \left( W[\Gamma_n] + \Delta F \right) } }
+ \sum\limits_{m=1}^{N_r} \frac{\mathcal{F}[\tilde{\Gamma}_m]}
{N_f + N_r e^{-\beta \left( W[\tilde{\Gamma}_m] + \Delta F \right) } } .
\end{eqnarray}
$N_f$ and $N_r$ are the number of trajectories sampled by the forward and the reverse process, respectively.

\subsection{Functionals for Equilibrium Properties}

While Equations \ref{eq:estimator_s}, \ref{eq:estimator_u}, and \ref{eq:estimator_b} may be used to estimate the path-ensemble average of any functional, specific choices of $\mathcal{F}[\Gamma]$ enable the calculation of equilibrium properties. We will consider two: $\Delta F_t$, the free energy difference between the system in the thermodynamic state at the beginning of the protocol and the thermodynamic state at time $t$; and $\Phi(z)$, the potential of mean force --- the free energy of a system as a function of a collective variable $z$. The quantities $\Delta F_t$ and $\Phi(z)$ differ in that the former includes the effect of an external biasing potential.

According to Jarzynski's equality \cite{Jarzynski1997,Jarzynski1997a},
\begin{eqnarray}
\label{eq:fe_estimate}
e^{- \beta \Delta F_t} = \E{e^{- \beta W[\Gamma(t)]}},
\end{eqnarray}
where $W[\Gamma(t)]$ is the work done up to time $t$. 
Therefore, $\Delta F_t$ may be estimated by using the functional $\mathcal{F}[\Gamma] = e^{-\beta W [\Gamma_n(t)] }$ in Equations \ref{eq:estimator_s}, \ref{eq:estimator_u}, and \ref{eq:estimator_b}.

To estimate $\Phi(z)$, we will use the probability density of $z$ in the thermodynamic state at time $t$, $p_t(z)$. This probability density may be obtained by \cite{Hummer2001,Hummer2005},
\begin{eqnarray}
p_t(z) = \E{ \delta \left[ z - z[\Gamma(t)] \right] e^{-\beta W [\Gamma(t)] } },
\end{eqnarray}
where $\delta$ is the delta function. Based on this equation, $p_t(z)$ may be estimated by using the functional $\mathcal{F}[\Gamma] = \delta \left[ z - z[\Gamma_n(t)] \right] e^{-\beta W [\Gamma_n(t)] }$ in Equations \ref{eq:estimator_s}, \ref{eq:estimator_u}, and \ref{eq:estimator_b}. 
We will focus on the situation where the total potential energy consists of a time-independent unbiased potential and a time-dependent harmonic bias on $z$. The harmonic bias may be described as, $V(z, t) = \frac{k}{2}[z - \lambda(t)]^2$, where $k$ is the spring constant, $z$ is the collective variable, and $\lambda(t)$ is the center of the harmonic potential at time $t$. In this situation, which describes a single-molecule pulling experiment, the probability density of $z$ at time $t$ can be readily reweighted into an unbiased counterpart. Then, $\hat{\Phi}(z)$ may be computed by combining $\hat{p}_t(z)$ for different $t$ \cite{Hummer2001,Hummer2005},
\begin{eqnarray}
\label{eq:pmf}
e^{- \beta \widehat{\Phi}(z)} =
\frac{\sum\limits_t \hat{p}_t(z) e^{\beta \Delta \hat{F}_t } }
{\sum\limits_t e^{-\beta \left[ V(z, t) - \Delta \hat{F}_t \right]}}.
\end{eqnarray}

\section{Computational Methods}

\subsection{System Setup}

\begin{figure}[p]
\includegraphics[scale=1]{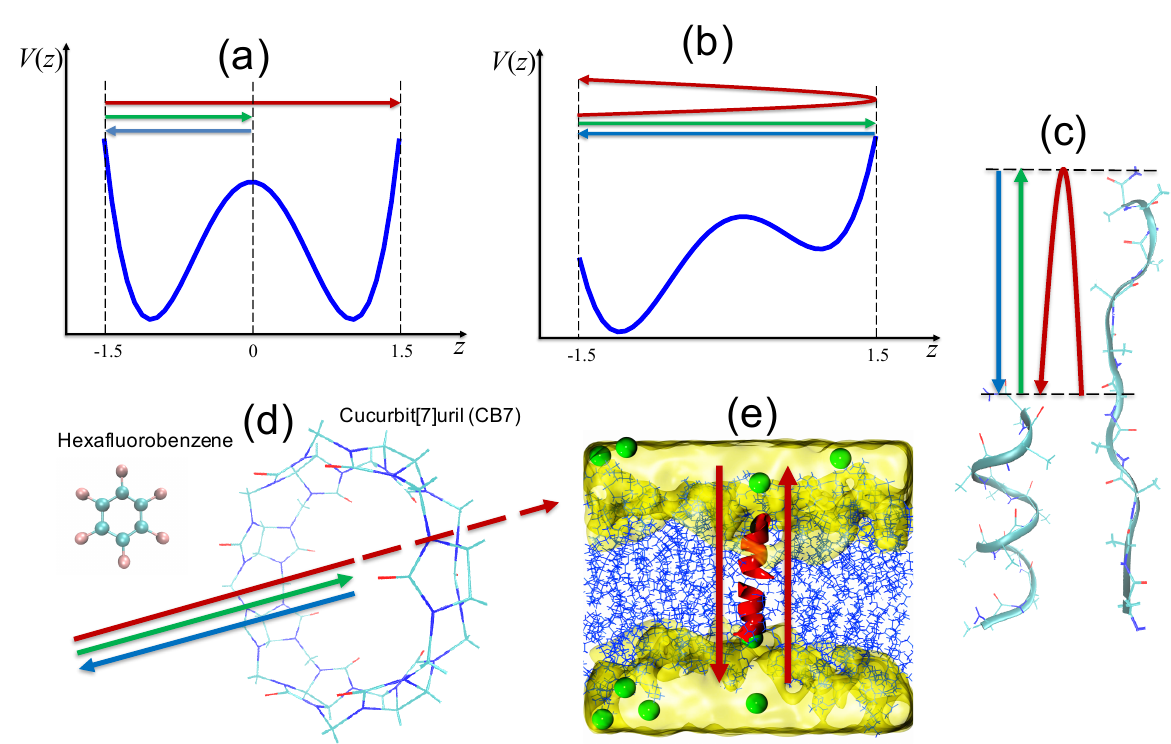}
\caption{
\label{fig:systems}
\textbf{Five systems considered in this paper.}
The systems are (a) a symmetric 1D potential, (b) symmetric 1D system, (c) deca-alanine, and (d) host-guest complex of CB7 and hexafluorobenzene, and (e) gramicidin A (red ribbon) inserted in a lipid bilayer (blue lines). In (e), water is represented by the yellow surface and potassium ions by green balls.
Forward (green), reverse (blue), and symmetric (red) processes are shown with colored arrows.
}
\end{figure}

Convergence properties of the statistical estimators were compared in five model systems: 
a symmetric (Figure \ref{fig:systems}a) and an asymmetric (Figure \ref{fig:systems}b) 1D potential, deca-alanine (Figure \ref{fig:systems}c), a host-guest complex of a cucurbit[7]uril (CB7) host and a hexafluorobenzene guest (Figure \ref{fig:systems}d), and a gramicidin A (gA) channel conducting the potassium ion (Figure \ref{fig:systems}e). 
The symmetric 1D potential was $U^{\textrm{sym}}(z) = 5(z^2 -1)^2$ (Figure \ref{fig:systems}a) and the asymmetric 1D potential was $U^{\textrm{asym}}(z) = (5z^3 - 10z + 3)z$ (Figure \ref{fig:systems}b). 
For deca-alanine, the structure (Figure \ref{fig:systems}c) and CHARMM22 force field \cite{Mackerell2004, MacKerell1998} parameters with the CMAP correction \cite{MacKerell2004b} for torsional angles in proteins were obtained from the NAMD \cite{Phillips2005} tutorial site \cite{Park2003,Park2012}.
The structure of the host molecule CB7 (Figure \ref{fig:systems}d) and AMBER force field parameters were generously provided by the Gilson group \cite{Velez-Vega2013}. (We also used this setup in a previous paper \cite{Nguyen2016}). In the provided parameters, partial atomic charges of CB7 were obtained from the VC/2004 parameter set \cite{Gilson2003} and Lennard-Jones and bonded parameters were based on AMBER99SB \cite{Hornak2006} and GAFF \cite{Wang2004a} force fields. The 3D structure of the guest molecule hexafluorobenzene was generated by BALLOON 1.5.0.1143 \cite{Vainio2007}. General AMBER force field (GAFF) parameters \cite{Wang2004a} from AMBER tools 16 and the AM1BCC partial charges from antechamber \cite{Wang2006} were used to parameterize the guest molecule. The host CB7 was oriented such that its symmetry axis was aligned along the $z$ axis and its center of mass was placed at the origin. The host's heavy atoms were fixed in all the simulations. 
The gA channel was set up in a previous study \cite{Ngo2016}. For completeness, some details are also described here. The gA channel pore was aligned along the $z$ coordinate and embedded in a lipid bilayer. The system was solvated in a water box with size 66.7 $\times$ 51.5 $\times$ 60.6 \AA$^3$ which also contained potassium and chloride ions (Figure \ref{fig:systems}e). The system was parameterized with the CHARMM 27 force field \cite{Mackerell2004, MacKerell1998}. CMAP (L-CMAP and inverted CMAP D-CMAP) corrections were used to model the phi-psi torsional angles present in the sequence of L- and D- amino-acid residues comprising each of the gA monomers.

\subsection{Nonequilibrium Pulling Simulations}

For the first four systems, we performed simulations with a forward, reverse, and symmetric process. For gA, we analyzed the symmetric simulations from \citet{Ngo2016}. 

For the 1D potential energy surfaces, Brownian dynamics simulations were performed using an in-house python script.
For the symmetric potential, the symmetric process consisted of pulling the Brownian particle at a constant speed from the equilibrated state at $\lambda=-1.5$ to $\lambda=1.5$ in 750 steps. In the forward asymmetric process, it was pulled from the equilibrated state at $\lambda=-1.5$ to $\lambda=0$. In the reverse process, it was pulled from the equilibrated state at $\lambda=0$ to $\lambda=-1.5$ in 375 steps (Figure \ref{fig:systems}a).
For the asymmetric potential, the symmetric process consisted of pulling the particle from the equilibrated state at $\lambda=-1.5$ to $\lambda=1.5$ and (without equilibrating) back to $\lambda=-1.5$ in 1500 steps. In the forward asymmetric process, it was pulled from the equilibrated state at $\lambda=-1.5$ to $\lambda=1.5$. In the reverse, it was pulled from the equilibrated state at $\lambda=1.5$ to $\lambda=-1.5$ in 750 steps (Figure \ref{fig:systems}b). The force constant $k$ of the pulling harmonic potential was chosen to be $k=15$ units of energy per one unit of length squared (for the 1D systems, distance, time, and energy have no specified unit).
For each system, we collected 20,000 pulling trajectories for the symmetric process and 40,000 for the asymmetric process.
Each set of trajectories was partitioned into 100 blocks.

For deca-alanine and the host-guest complex, Langevin dynamics simulations were performed using NAMD 2.9 \cite{Phillips2005}. The systems were in vacuum. Bond distances involving hydrogen atoms were constrained using the SHAKE algorithm \cite{Smith1994} which enabled the use of a time step of 2 fs. Nonbonded cutoffs were set to 999 \AA. At each end state the structures were minimized for 1000 steps and the temperature was increased by 10 K every 100 steps from 0 to 300 K. After discarding the first 100 ps, equilibrated structures were collected every 1 ps over a total production time of 1 ns at at 300 K.

With deca-alanine, one end of the peptide was fixed and the other end was attached to the harmonic potential $V(d, t)$, where $d$ is the distance between the two ends. The force constant was set to $k=7.2$ kcal/mol/\AA$^2$, which is the same as in a previous paper \cite{Park2003}.
The peptide was equilibrated at both $\lambda = 1.3$ nm and $\lambda = 3.3$ nm for 1 ns. 
In the symmetric process, the spring was pulled from the equilibrated state at $\lambda = 1.3$ nm to 3.3 nm and, without equilibrating, compressed it back to 1.3 nm. 
In the forward asymmetric process, it was pulled from the equilibrated state at $\lambda = 1.3$ nm to 3.3 nm, while in the reverse counterpart it was pulled from the equilibrated state at $\lambda = 3.3$ nm to 1.3 nm. The pulling speed for all the processes was set to 2 nm per 1 ns.
We collected 200, 400, and 400 trajectories for the symmetric, forward, and reverse asymmetric processes, respectively. 
Each set of trajectories was partitioned into 10 blocks.

With the host-guest complex, the harmonic pulling potential $V(z, t)$ was attached to the center of mass of the guest molecule and $\lambda$ was set to the $z$ coordinate of harmonic trap position. The spring constant was chosen as $k=7.2$ kcal/mol/\AA$^2$, which is the same as for deca-alanine.
The complex was equilibrated at both $\lambda = -2$ nm and  $\lambda = 0$ nm for 1 ns. In the symmetric process, the spring pulled the guest molecule from the equilibrated state at $\lambda = -2$ nm through the pore of the host to $\lambda = 2$ nm. In the forward asymmetric process, it was pulled from the equilibrated state at $\lambda = -2$ nm to the center of the host at $\lambda = 0$ nm. In its reverse counterpart, the spring was pulled from the equilibrated state at $\lambda = 0$ nm to $\lambda = -2$ nm (Figure \ref{fig:systems}d). The pulling speed was the same for all processes and set to 2.5 nm per ns. We collected 400, 800 and 800 trajectories for the symmetric, forward and reverse processes, respectively.
Each set of trajectories was partitioned into 10 blocks.

In the pulling simulations described in \citet{Ngo2016}, the potassium ion was pulled from $z=-1.3$ nm to $z=1.3$ nm using a harmonic force constant of $k=100$ kcal/mol/\AA$^2$. Two pulling speeds were used, $v=100$ \AA/ns and $v=10$ \AA/ns. For the fast pulling simulations, 145 forward (pulling from $z=-1.3$ nm to $z=1.3$ nm) and 145 reverse (from $z=1.3$ nm to $z=-1.3$ nm) trajectories were collected. For the slow pulling simulations, the number of trajectories collected were 16 for forward and 16 for reverse pulling. 

\subsection{Reference free energies and PMFs}

For the 1D potentials, reference values were based on numerical integration and from the exact PMF. The exact PMF is simply the potential itself. The reference free energy was computed by numerically integrating $F_t = \int e^ {-\beta \left[ U(z) + V(z, t)\right] }dz$ between $\lambda(t)-5 < z < \lambda(t)+5$, where $U(z)$ is $U^{\textrm{sym}}(z)$ or $U^{\textrm{asym}}(z)$, using the adaptive quadrature method implemented in SciPy 0.17.0.

For the other systems, reference free energies and PMFs were based on umbrella sampling and WHAM. For deca-alanine, umbrella sampling was based on 100 windows which were equally spaced between $\lambda = 1.3$ nm and $\lambda = 3.3$ nm. For the host-guest complex, we used 200 windows equally spaced between $\lambda = -2$ nm and $\lambda = 2$ nm. In each window, we equilibrated for 1 ns and ran production simulation for 20 ns. The spring constant was set to $k=7.2$ kcal/mol/\AA$^2$, which is the same as in the nonequilibrium pulling simulations. As in the pulling processes, Langevin dynamics simulations at 300 K of each system in vacuum were performed using NAMD 2.9 \cite{Phillips2005} with a time step of 2 fs. In the gA system, umbrella sampling was carried out in \citet{Ngo2016} using 49 windows equally spaced between $z=-1.2$ nm and $z=1.2$ nm.

We used the multistate Bennett acceptance ratio \cite{Shirts2008} (MBAR) to estimate the free energy differences and the weighted histogram analysis method \cite{Ferrenberg1989,Kumar1992} (WHAM) to estimate PMFs.

\subsection{Analysis}

In the first four systems, the nonequilibrium work data generated by three processes --- forward (f), reverse (r), and symmetric (s) --- was combined with three estimators --- unidirectional (u), bidirectional (b) and symmetric (s) --- to produce six different estimates of the free energy and PMF. For symmetric processes, the generated data were used with all three estimators to give three estimates: s\_u, s\_b and s\_s. When using the bidirectional estimator, we split the total set of data into two halves, considering the first half as forward and the second as reverse because the symmetric process is both forward and reverse. For asymmetric (forward and reverse) processes, we applied the unidirectional estimator to each direction to produce f\_u and r\_u estimates and combined both directions using the bidirectional estimator to obtain the fr\_b estimate. When asymmetric processes were used in symmetric systems, we exploited the symmetry of the system by replicating free energies and PMFs from one half of the process or system to the other half. To make the comparison of symmetric and asymmetric processes as fair as possible, the fact that asymmetric processes were half the length was compensated for by analyzing twice as many trajectories. For every system, free energy and PMF estimates were generated for each of the 10 or 100 blocks. The estimates were used to calculate the RMSE with respect to reference values.

In gA, nonequilibrium work data were processed with the unidirectional (u), bidirectional (b), symmetric (s) estimators as well as b+WHAM estimator \cite{Ngo2016}, which uses bidirectional data to initiate the weighted histogram analysis method.

\section{Results}

\subsection{1D systems}

\begin{figure}[p]
\vspace{-15pt}
\includegraphics[scale=1]{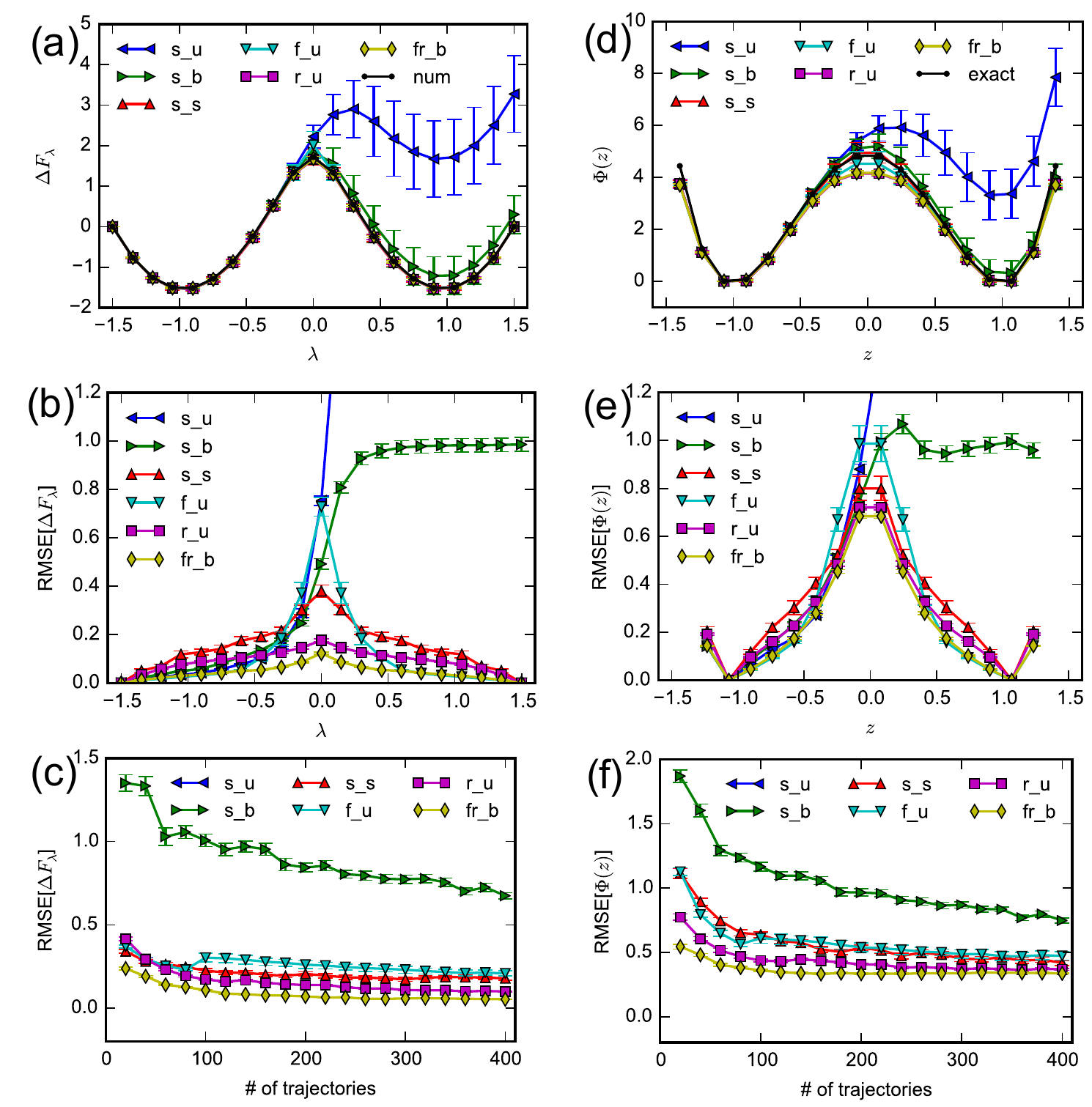}
\vspace{-15pt}
\caption{
\label{fig:1d_symm_fe_pmf}
\textbf{Free energies and PMFs for the symmetric 1D system.} 
Lines are labeled by combining the name of the nonequilibrium process: ``s'' for symmetric process, ``f'' for forward, and ``r'' for reverse asymmetric process, with name of the estimator: ``u'' for unidirectional, ``b'' for bidirectional (MA) and ``s'' for symmetric.
[(a) and (d)]: Comparison of free energies and PMFs estimated by nonequilibrium path ensemble average estimators with numerical free energy (num) and exact PMF (exact).
[(b) and (e)]: RMSE of free energies and PMFs estimated by non-equilibrium path ensemble average estimators with respect to the numerical free energies and exact PMF, respectively.
[(c) and (f)]: Convergence of RMSE of free energies and PMFs with the number of non-equilibrium trajectories. RMSEs are averaged over $\lambda$ for free energies or bins for PMFs. For f\_u, r\_u and fr\_u, the actual number of trajectories is double the number shown in panels (c) and (f).
In panels (b) and (e), RMSEs greater than 1.2 are not shown.
In panels (c) and (f), s\_u lines are larger than 1.5 and 2, respectively, and are not shown.
}
\end{figure}

For the symmetric potential, if work data is generated by a symmetric process, then applying the our new symmetric estimator, Equation \ref{eq:estimator_s}, gives more accurate free energy and PMF estimates than the previous unidirectional (Equation \ref{eq:estimator_u}) and bidirectional (Equation \ref{eq:estimator_b}) estimators.
Free energy and PMF estimates from the unidirectional and bidirectional estimators, s\_u and s\_b, respectively, show large biases in the right half of the potential (Figure \ref{fig:1d_symm_fe_pmf}). 
The bias in the unidirectional estimator is substantially larger. 
When using the symmetric estimator (s\_s), RMSEs of the free energies and PMFs are significantly reduced.

Although the symmetric estimator leads to the best performance when analyzing data from symmetric processes, better performance is achieved with data from asymmetric processes. The lowest RMSEs for both free energies and PMFs are obtained when applying the bidirectional estimator to asymmetric processes. When using the unidirectional estimator, the RMSEs depend quite significantly on the direction of pulling, e.g. forward from $\lambda = -1.5$ to 0 or reverse from $\lambda = 0$ to -1.5. For this system, r\textunderscore u gives much lower RMSEs than f\_u, especially around the barrier region of the 1D potential (Figure \ref{fig:1d_symm_fe_pmf}). Free energy and PMF estimates from our new symmetric estimator s\_s have lower RMSEs than r\_u but larger RMSEs than r\_u and fr\_b, although the difference is less significant for the PMFs.

The fr\_b estimates show the fastest convergence with respect to the number of trajectories (Figures \ref{fig:1d_symm_fe_pmf}c and f). The convergence of our new estimator s\_s is slightly slower than r\_u and faster than f\_u. Both s\_u and s\_b show extremely slow convergence (s\_u is not shown because it is out of range).

\begin{figure}[p]
\vspace{-15pt}
\includegraphics[scale=1]{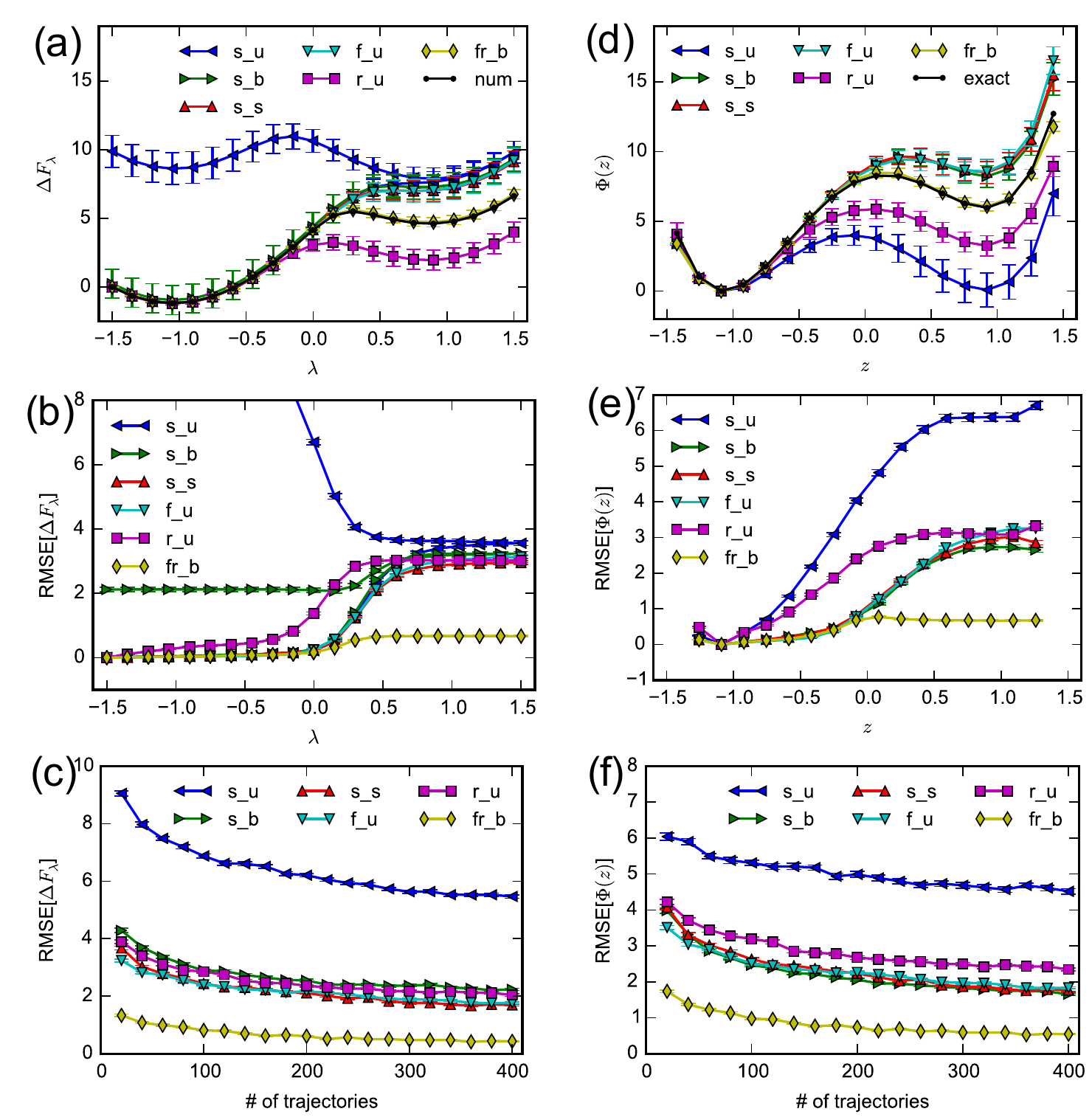}
\vspace{-15pt}
\caption{
\label{fig:1d_asymm_fe_pmf}
\textbf{Free energies and PMFs for the asymmetric 1D system.} 
Lines are labeled by combining the name of the nonequilibrium process: ``s'' for symmetric process, ``f'' for forward, and ``r'' for reverse asymmetric process, with name of the estimator: ``u'' for unidirectional, ``b'' for bidirectional (MA) and ``s'' for symmetric.
[(a) and (d)]: Comparison of free energies and PMFs estimated by nonequilibrium path ensemble average estimators with numerical free energy (num) and exact PMF (exact).
[(b) and (e)]: RMSE of free energies and PMFs estimated by non-equilibrium path ensemble average estimators with respect to the numerical free energies and exact PMF, respectively.
[(c) and (f)]: Convergence of RMSE of free energies and PMFs with the number of non-equilibrium trajectories. RMSEs are averaged over $\lambda$ for free energies or bins for PMFs. For f\_u, r\_u and fr\_u, the actual number of trajectories is double the number shown in panels (c) and (f).
In (b), free energy RMSEs larger than 8 are not shown.}
\end{figure}

Similar trends hold for the asymmetric 1D potential. 
s\_s is the best of estimators for symmetric processes, but fr\_b has the best performance overall. 
For all estimators except fr\_b, there is significant bias in the region to the right of $z=0$. 
s\_u estimates of free energies and PMFs have the largest RMSEs (Figure \ref{fig:1d_asymm_fe_pmf}). 
Unlike the symmetric system above, the bidirectional estimator combined with symmetric process (s\_b) gives almost identical performance to our new symmetric estimator (s\_s). 
The forward asymmetric process from $\lambda = -1.5$ to $\lambda = 1.5$ (f\_u) is better than the reverse counterpart (r\_u) and is nearly identical to s\_s.
fr\_b estimates show the fastest convergence with respect to the number of trajectories. All other estimators except for s\_u, which is particularly slow, converge at a similar rate.

\subsection{Deca-alanine}

\begin{figure}[p]
\vspace{-15pt}
\includegraphics[scale=1]{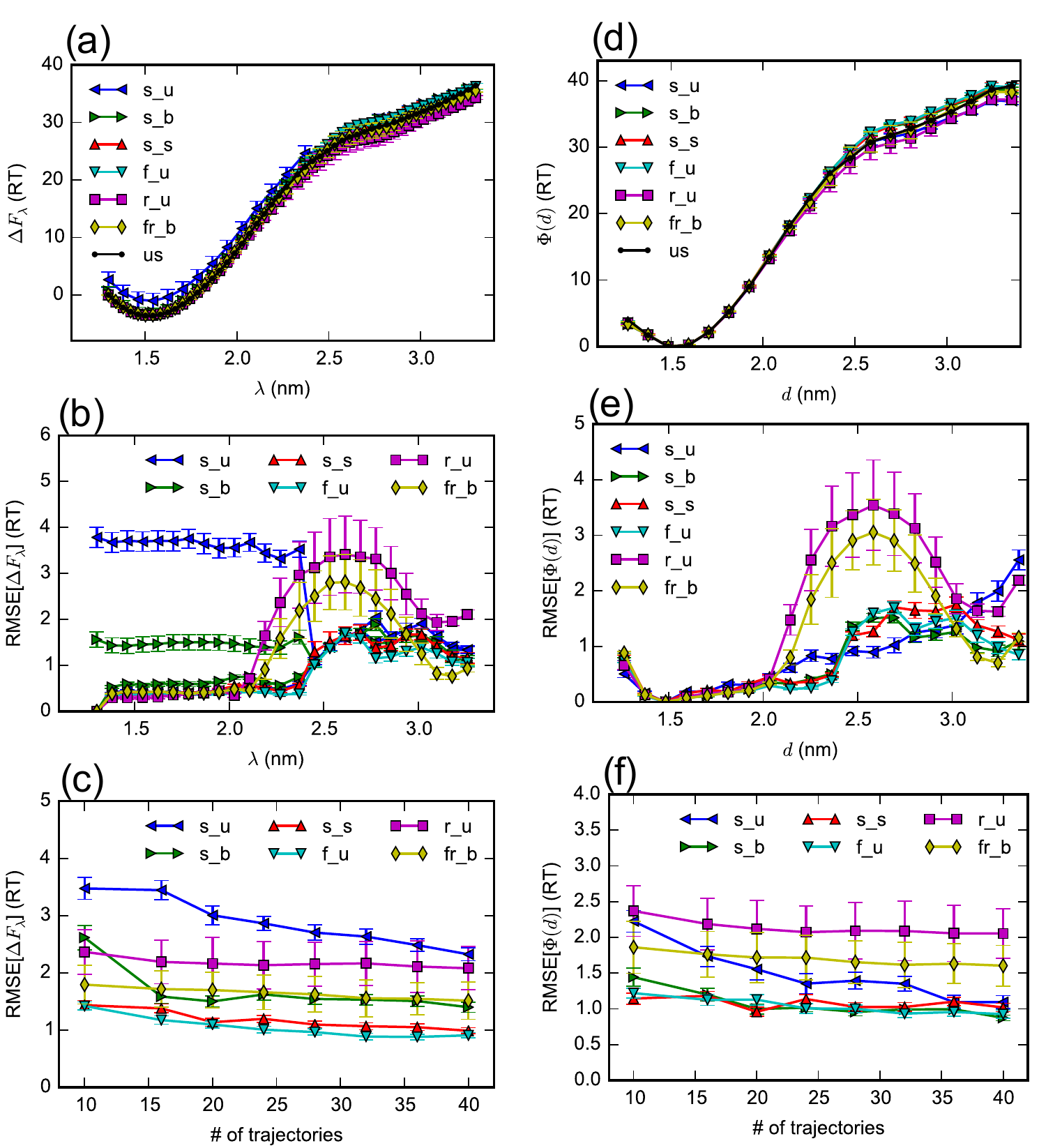}
\vspace{-15pt}
\caption{
\label{fig:deca_fe_pmf}
\textbf{Free energies and PMFs for deca-alanine.} 
Lines are labeled by combining the name of the nonequilibrium process: ``s'' for symmetric process, ``f'' for forward, and ``r'' for reverse asymmetric process, with name of the estimator: ``u'' for unidirectional, ``b'' for bidirectional (MA) and ``s'' for symmetric.
[(a) and (d)]: Comparison of free energies and PMFs estimated by nonequilibrium path ensemble average estimators with numerical free energy (num) and exact PMF (exact).
[(b) and (e)]: RMSE of free energies and PMFs estimated by non-equilibrium path ensemble average estimators with respect to the numerical free energies and exact PMF, respectively.
[(c) and (f)]: Convergence of RMSE of free energies and PMFs with the number of non-equilibrium trajectories. RMSEs are averaged over $\lambda$ for free energies or bins for PMFs. For f\_u, r\_u and fr\_u, the actual number of trajectories is double the number shown in panels (c) and (f).
}
\end{figure}

The performance of estimators for deca-alanine is distinct from the 1D potentials 
(Figure \ref{fig:deca_fe_pmf}).
In particular, fr\_b has especially weak performance, especially when $\lambda$ or $d$ are near 2 nm.
In its place, s\_s and f\_u are the best estimators, with comparably low RMSEs and convergence properties for both free energies and PMFs.
Remarkably, s\_u, which shows the worst performance for the 1D systems, gives rather low RMSE for the PMFs (Figure \ref{fig:deca_fe_pmf}) and the RMSE decreases dramatically with the number of trajectories (Figures \ref{fig:deca_fe_pmf}c and f). Among the estimates arising from the asymmetric processes, f\_u gives the lowest RMSEs for both free energies and PMFs whereas r\_u and fr\_b give significantly larger RMSEs. In this case, the reverse pulling is extremely unhelpful and combining it with the forward counterpart using the bidirectional estimator does not improve free energy estimates.

\subsection{Host-guest complex}

\begin{figure}[p]
\vspace{-15pt}
\includegraphics[scale=1]{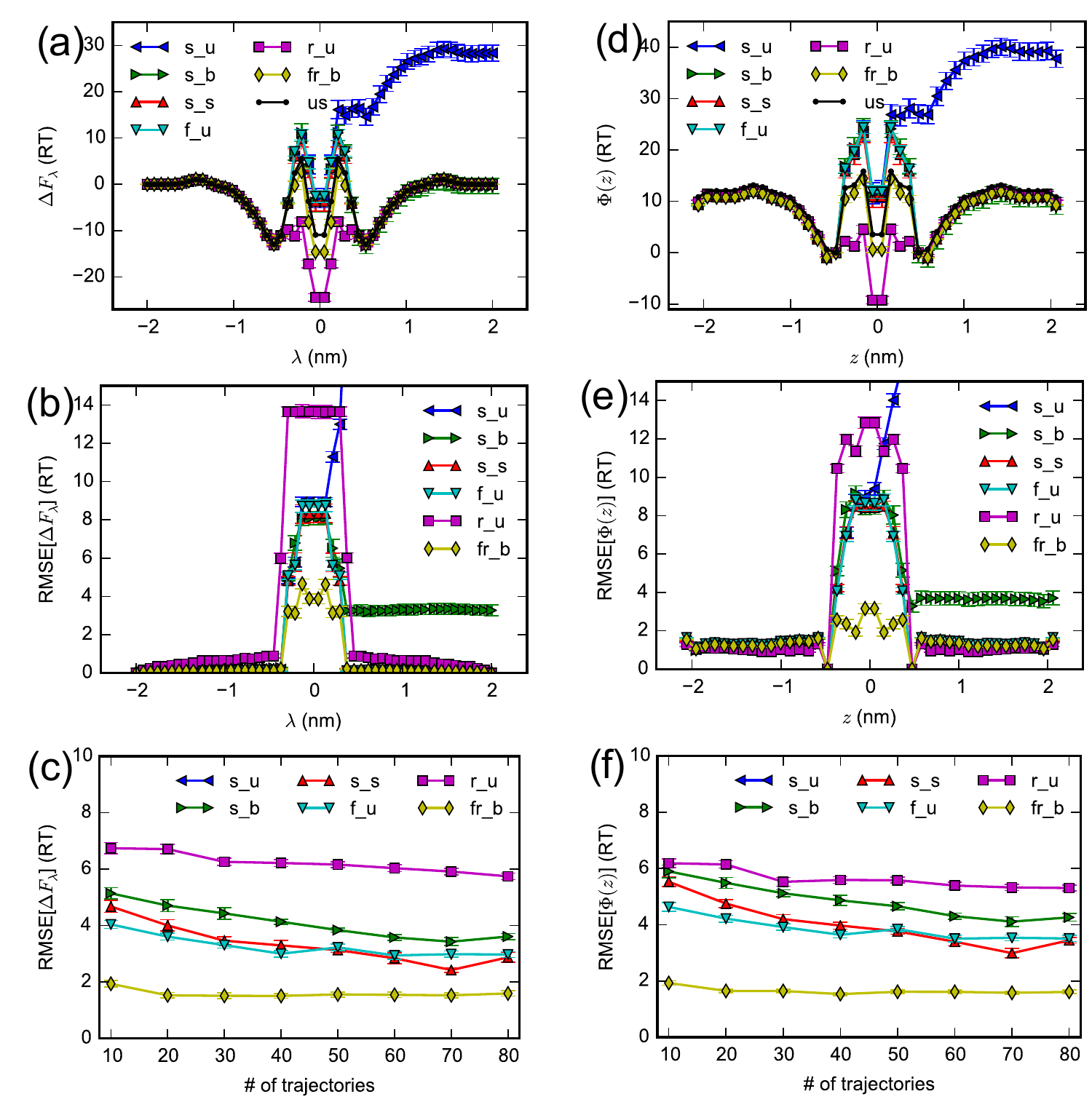}
\vspace{-15pt}
\caption{
\label{fig:cuc7_fe_pmf}
\textbf{Free energies and PMFs for host-guest complex.} 
Lines are labeled by combining the name of the nonequilibrium process: ``s'' for symmetric process, ``f'' for forward, and ``r'' for reverse asymmetric process, with name of the estimator: ``u'' for unidirectional, ``b'' for bidirectional (MA) and ``s'' for symmetric.
[(a) and (d)]: Comparison of free energies and PMFs estimated by nonequilibrium path ensemble average estimators with numerical free energy (num) and exact PMF (exact).
[(b) and (e)]: RMSE of free energies and PMFs estimated by non-equilibrium path ensemble average estimators with respect to the numerical free energies and exact PMF, respectively.
[(c) and (f)]: Convergence of RMSE of free energies and PMFs with the number of non-equilibrium trajectories. RMSEs are averaged over $\lambda$ for free energies or bins for PMFs. For f\_u, r\_u and fr\_u, the actual number of trajectories is double the number shown in panels (c) and (f).
In panels (b) and (e), RMSEs larger than 15 RT are not shown.}
\end{figure}

Free energy estimates in the host-guest complex are notable because of much larger error inside versus outside of the host (Figure \ref{fig:cuc7_fe_pmf}). As in most systems, s\_s is the best of estimates based on symmetric protocols and fr\_b has the lowest RMSE overall. f\_u has a similar performance to s\_s. The s\_u estimates of free energies and PMFs again have the largest RMSEs, especially on the right half of the free energy and PMF. In all estimators except for r\_u and fr\_b, free energies and PMFs are overestimated inside of the host. In fr\_b, the quantities are slightly underestimated. In r\_u, the underestimation is much more significant.

\subsection{Gramicidin A}

\begin{figure}[p]
\includegraphics[scale=1]{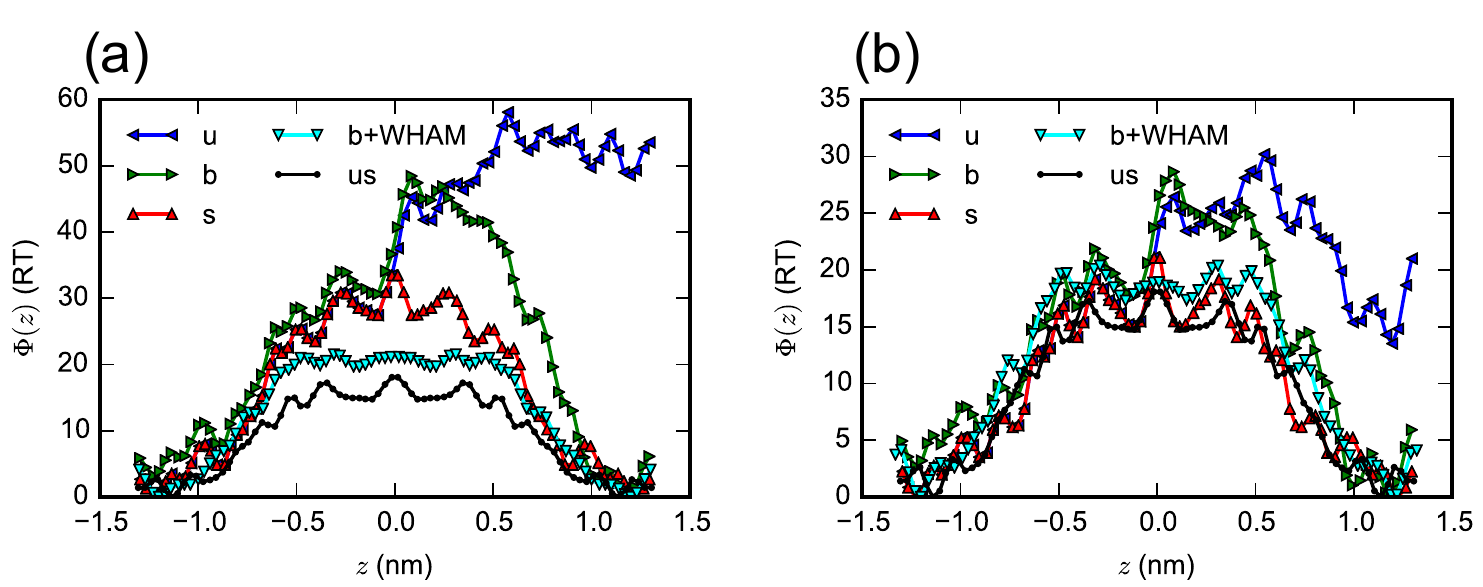}
\caption{
\label{fig:gA_pmf}
\textbf{PMFs for gA.}
The pulling speed is (a) $v=100$ \AA/ns and (b) $v=10$ \AA/ns.
Comparison of PMFs from unidirectional (u), bidirectional (b), symmetric (s), combined bidirectional and WHAM (b+WHAM), and umbrella sampling (us). 
}
\end{figure}

Depending on the pulling speed, Equation \ref{eq:estimator_s} (s) leads to either the most accurate or second most accurate PMF estimates. For the fast pulling speed, the b+WHAM estimate is the least biased (Figure \ref{fig:gA_pmf}a). Equation \ref{eq:estimator_s} (s) is less accurate than b+WHAM, with the largest bias occurring near the energy barrier. On the other hand, Equation \ref{eq:estimator_u} (u) and Equation \ref{eq:estimator_b} (b) show significant bias, especially on the right side of the energy barrier. When the pulling is slow (Figure \ref{fig:gA_pmf}b), Equation \ref{eq:estimator_s} (s) leads to the most accurate PMF estimates, most accurately reproducing the shape of the barrier. On the other hand, b+WHAM is slightly biased. As with the fast pulling, Equation \ref{eq:estimator_u} (u) and Equation \ref{eq:estimator_b} (b) are highly biased on the right side, when $z > 0$.

\section{Discussion}

Demonstrative calculations on 1D potentials and molecular systems show that if work data are generated by a symmetric process, it is always best to use our newly derived symmetric estimator (s\_s). The symmetric estimator gives lower RMSEs compared to unidirectional (s\_u) and bidirectional (s\_b) estimators. On the other hand, if they are generated from a pair of asymmetric processes, the bidirectional estimator (fr\_b) usually gives the lowest overall RMSEs. 

The host-guest complex illustrates the benefits of using a bidirectional estimator. The trajectories of nonequilibrium driven processes that are most useful for reconstructing equilibrium properties are close to equilibrium. When using only the forward process of inserting the guest into the host (f\_u), the barrier to insertion is accurately determined but the depth of the well in the bound complex, which is farthest from the initial state, is not low enough. On the other hand, when only using the reverse process of extracting the guest (r\_u), the bound state is close to equilibrium and the well depth appears to be accurate. When the complex starts to dissociate, however, the system fails to form interactions that would reduce the work and lead to a smaller barrier to extraction, so the overall well depth is too low.

Although bidirectional estimation generally has better performance, there appears to be an exception if one process is much further from equilibrium than its counterpart, as in deca-alanine. In this system, the forward process involves unfolding of the peptide and the reverse process corresponds to refolding a helix from an extended state. Due to a large number of structural rearrangements required for refolding, the system is unlikely to find the native conformation during reverse process. Hence, estimates from the reverse process are biased and r\_u has a large RMSE. Due to the relatively far-from-equilibrium nature of reverse trajectories, the bidirectional estimator has worse performance than f\_u.

In addition to cases where one direction is further from equilibrium than its counterpart, the bidirectional estimator has the disadvantage that the system needs to be equilibrated at both ends. This may not always be feasible. In a simulation, equilibration consumes more computer time. In a single-molecule pulling experiment, waiting for equilibration can accumulate more drift.

In situations where a bidirectional process is not feasible or beneficial, a symmetric protocol and Equation \ref{eq:estimator_s} is an attractive alternative to a unidirectional process and analysis; it can significantly reduce the experimental measurement or computer simulation time required to attain the same accuracy. For example, in a single-molecule pulling experiment, a forward protocol that unfolds a molecule can be immediately followed by its time reversal to create a symmetric protocol. As the performance of s\_s and f\_u for the same amount of simulation time are remarkably similar, our results suggest that using a symmetric protocol to return to the initial position of the apparatus before equilibration is as valuable as collecting data from twice as many forward processes alone. In contrast to using only forward processes, however, the symmetric protocol reduces the time in equilibration and makes use of data from the return to the initial apparatus position. As another example, simulating the pulling of a small organic molecule completely across a symmetric lipid membrane and using Equation \ref{eq:estimator_s} could be useful for membrane permeability studies. In this case, it can be significantly more difficult to equilibrate the molecule in the middle of the membrane than in the solvent. If the small molecule is pulled completely through the membrane using a symmetric protocol, our data suggest that Equation \ref{eq:estimator_s} will yield more accurate results than the unidirectional estimator, Equation \ref{eq:estimator_u}.

There are several tradeoffs in the choice between b+WHAM and the symmetric estimator, Equation \ref{eq:estimator_s}.
On one hand, gA simulations suggest that b+WHAM is more capable of reducing bias when estimating equilibrium properties using far-from-equilibrium trajectories.
On the other hand, the symmetric estimator is more theoretically rigorous because it does not assume that configurations are drawn from equilibrium distributions (an assumption in WHAM).
Consequently, it appears to reconstruct the shape of PMFs more accurately.
Figure \ref{fig:gA_pmf} also highlights the importance of selecting an appropriate pulling speed. Regardless of the estimator, PMFs reconstructed from the processes that are closer to equilibrium are more accurate than those from the faster pullings.

\section{Conclusions}

We have developed a new and theoretically rigorous statistical estimator for estimating nonequilbrium path-ensemble averages for driven processes with symmetric protocols. In demonstrative calculations, the symmetric estimator outperforms existing rigorous estimators when using such trajectories to estimate free energies and PMFs. If bidirectional trajectories initiated from two equilibrated states are available, however, they usually outperform the symmetric estimator. Hence, we only recommend using symmetric protocols and the new estimator in situations where the bidirectional approach is not practical or expected to be beneficial. Results from the symmetric estimator are remarkably similar to those from twice as many forward process that constitute the first half of the symmetric protocol.


\begin{acknowledgments}

We thank high summer intern John Zaris for his effort in the early stages of this project and undergraduate summer intern Luiz Matheus Barbosa Santos for doing some work with Jo\~{a}o. Jo\~{a}o and Luiz were funded by the Capes Foundation within the Brazilian Ministry of Education. Van Ngo is supported by a LANL Director's Fellowship (2018-2021). The work in Calgary was supported by the Natural Sciences and Engineering Research Council of Canada (NSERC) (Discovery Grant RGPIN-315019 to SYN) and the Alberta Innovates Technology Futures (AITF) Strategic Chair in BioMolecular Simulations (Centre for Molecular Simulation). Molecular simulations were performed under Resource Allocation Award by Compute Canada (2018-19).

\end{acknowledgments}



\begin{thebibliography}{45}%
\makeatletter
\providecommand \@ifxundefined [1]{%
 \@ifx{#1\undefined}
}%
\providecommand \@ifnum [1]{%
 \ifnum #1\expandafter \@firstoftwo
 \else \expandafter \@secondoftwo
 \fi
}%
\providecommand \@ifx [1]{%
 \ifx #1\expandafter \@firstoftwo
 \else \expandafter \@secondoftwo
 \fi
}%
\providecommand \natexlab [1]{#1}%
\providecommand \enquote  [1]{``#1''}%
\providecommand \bibnamefont  [1]{#1}%
\providecommand \bibfnamefont [1]{#1}%
\providecommand \citenamefont [1]{#1}%
\providecommand \href@noop [0]{\@secondoftwo}%
\providecommand \href [0]{\begingroup \@sanitize@url \@href}%
\providecommand \@href[1]{\@@startlink{#1}\@@href}%
\providecommand \@@href[1]{\endgroup#1\@@endlink}%
\providecommand \@sanitize@url [0]{\catcode `\\12\catcode `\$12\catcode
  `\&12\catcode `\#12\catcode `\^12\catcode `\_12\catcode `\%12\relax}%
\providecommand \@@startlink[1]{}%
\providecommand \@@endlink[0]{}%
\providecommand \url  [0]{\begingroup\@sanitize@url \@url }%
\providecommand \@url [1]{\endgroup\@href {#1}{\urlprefix }}%
\providecommand \urlprefix  [0]{URL }%
\providecommand \Eprint [0]{\href }%
\providecommand \doibase [0]{http://dx.doi.org/}%
\providecommand \selectlanguage [0]{\@gobble}%
\providecommand \bibinfo  [0]{\@secondoftwo}%
\providecommand \bibfield  [0]{\@secondoftwo}%
\providecommand \translation [1]{[#1]}%
\providecommand \BibitemOpen [0]{}%
\providecommand \bibitemStop [0]{}%
\providecommand \bibitemNoStop [0]{.\EOS\space}%
\providecommand \EOS [0]{\spacefactor3000\relax}%
\providecommand \BibitemShut  [1]{\csname bibitem#1\endcsname}%
\let\auto@bib@innerbib\@empty
\bibitem [{\citenamefont {Jarzynski}(1997{\natexlab{a}})}]{Jarzynski1997}%
  \BibitemOpen
  \bibfield  {author} {\bibinfo {author} {\bibfnamefont {C.}~\bibnamefont
  {Jarzynski}},\ }\href@noop {} {\bibfield  {journal} {\bibinfo  {journal}
  {Phys. Rev. E}\ }\textbf {\bibinfo {volume} {56}},\ \bibinfo {pages} {5018}
  (\bibinfo {year} {1997}{\natexlab{a}})}\BibitemShut {NoStop}%
\bibitem [{\citenamefont {Jarzynski}(1997{\natexlab{b}})}]{Jarzynski1997a}%
  \BibitemOpen
  \bibfield  {author} {\bibinfo {author} {\bibfnamefont {C.}~\bibnamefont
  {Jarzynski}},\ }\href@noop {} {\bibfield  {journal} {\bibinfo  {journal}
  {Phys. Rev. Lett.}\ }\textbf {\bibinfo {volume} {78}},\ \bibinfo {pages}
  {2690} (\bibinfo {year} {1997}{\natexlab{b}})}\BibitemShut {NoStop}%
\bibitem [{\citenamefont {Crooks}(2000)}]{Crooks2000}%
  \BibitemOpen
  \bibfield  {author} {\bibinfo {author} {\bibfnamefont {G.~E.}\ \bibnamefont
  {Crooks}},\ }\href@noop {} {\bibfield  {journal} {\bibinfo  {journal} {Phys.
  Rev. E}\ }\textbf {\bibinfo {volume} {61}},\ \bibinfo {pages} {2361}
  (\bibinfo {year} {2000})}\BibitemShut {NoStop}%
\bibitem [{\citenamefont {Hummer}\ and\ \citenamefont
  {Szabo}(2001)}]{Hummer2001}%
  \BibitemOpen
  \bibfield  {author} {\bibinfo {author} {\bibfnamefont {G.}~\bibnamefont
  {Hummer}}\ and\ \bibinfo {author} {\bibfnamefont {A.}~\bibnamefont {Szabo}},\
  }\href@noop {} {\bibfield  {journal} {\bibinfo  {journal} {Proc. Natl. Acad.
  Sci. USA}\ }\textbf {\bibinfo {volume} {98}},\ \bibinfo {pages} {3658}
  (\bibinfo {year} {2001})}\BibitemShut {NoStop}%
\bibitem [{\citenamefont {Hummer}\ and\ \citenamefont
  {Szabo}(2005)}]{Hummer2005}%
  \BibitemOpen
  \bibfield  {author} {\bibinfo {author} {\bibfnamefont {G.}~\bibnamefont
  {Hummer}}\ and\ \bibinfo {author} {\bibfnamefont {A.}~\bibnamefont {Szabo}},\
  }\href@noop {} {\bibfield  {journal} {\bibinfo  {journal} {Acc. Chem. Res.}\
  }\textbf {\bibinfo {volume} {38}},\ \bibinfo {pages} {504} (\bibinfo {year}
  {2005})}\BibitemShut {NoStop}%
\bibitem [{\citenamefont {Minh}(2006)}]{Minh2006a}%
  \BibitemOpen
  \bibfield  {author} {\bibinfo {author} {\bibfnamefont {D.~D.~L.}\
  \bibnamefont {Minh}},\ }\href@noop {} {\bibfield  {journal} {\bibinfo
  {journal} {Phys. Rev. E}\ }\textbf {\bibinfo {volume} {74}},\ \bibinfo
  {pages} {61120} (\bibinfo {year} {2006})}\BibitemShut {NoStop}%
\bibitem [{\citenamefont {Minh}(2007)}]{Minh2007a}%
  \BibitemOpen
  \bibfield  {author} {\bibinfo {author} {\bibfnamefont {D.~D.~L.}\
  \bibnamefont {Minh}},\ }\href@noop {} {\bibfield  {journal} {\bibinfo
  {journal} {J. Phys. Chem. B}\ }\textbf {\bibinfo {volume} {111}},\ \bibinfo
  {pages} {4137} (\bibinfo {year} {2007})}\BibitemShut {NoStop}%
\bibitem [{\citenamefont {Minh}\ and\ \citenamefont
  {McCammon}(2008)}]{Minh2008a}%
  \BibitemOpen
  \bibfield  {author} {\bibinfo {author} {\bibfnamefont {D.~D.~L.}\
  \bibnamefont {Minh}}\ and\ \bibinfo {author} {\bibfnamefont {J.~A.}\
  \bibnamefont {McCammon}},\ }\href@noop {} {\bibfield  {journal} {\bibinfo
  {journal} {J. Phys. Chem. B}\ }\textbf {\bibinfo {volume} {112}},\ \bibinfo
  {pages} {5892} (\bibinfo {year} {2008})}\BibitemShut {NoStop}%
\bibitem [{\citenamefont {Minh}\ and\ \citenamefont {Adib}(2008)}]{Minh2008}%
  \BibitemOpen
  \bibfield  {author} {\bibinfo {author} {\bibfnamefont {D.~D.~L.}\
  \bibnamefont {Minh}}\ and\ \bibinfo {author} {\bibfnamefont {A.~B.}\
  \bibnamefont {Adib}},\ }\href@noop {} {\bibfield  {journal} {\bibinfo
  {journal} {Phys. Rev. Lett.}\ }\textbf {\bibinfo {volume} {100}},\ \bibinfo
  {pages} {180602} (\bibinfo {year} {2008})}\BibitemShut {NoStop}%
\bibitem [{\citenamefont {Minh}\ and\ \citenamefont
  {Chodera}(2009)}]{Minh2009}%
  \BibitemOpen
  \bibfield  {author} {\bibinfo {author} {\bibfnamefont {D.~D.~L.}\
  \bibnamefont {Minh}}\ and\ \bibinfo {author} {\bibfnamefont {J.~D.}\
  \bibnamefont {Chodera}},\ }\href@noop {} {\bibfield  {journal} {\bibinfo
  {journal} {J. Chem. Phys.}\ }\textbf {\bibinfo {volume} {131}},\ \bibinfo
  {pages} {134110} (\bibinfo {year} {2009})}\BibitemShut {NoStop}%
\bibitem [{\citenamefont {Hummer}\ and\ \citenamefont
  {Szabo}(2010)}]{Hummer2010}%
  \BibitemOpen
  \bibfield  {author} {\bibinfo {author} {\bibfnamefont {G.}~\bibnamefont
  {Hummer}}\ and\ \bibinfo {author} {\bibfnamefont {A.}~\bibnamefont {Szabo}},\
  }\href@noop {} {\bibfield  {journal} {\bibinfo  {journal} {Proc. Natl. Acad.
  Sci. USA}\ }\textbf {\bibinfo {volume} {107}},\ \bibinfo {pages} {21441}
  (\bibinfo {year} {2010})}\BibitemShut {NoStop}%
\bibitem [{\citenamefont {Park}\ \emph {et~al.}(2003)\citenamefont {Park},
  \citenamefont {Khalili-Araghi}, \citenamefont {Tajkhorshid},\ and\
  \citenamefont {Schulten}}]{Park2003}%
  \BibitemOpen
  \bibfield  {author} {\bibinfo {author} {\bibfnamefont {S.}~\bibnamefont
  {Park}}, \bibinfo {author} {\bibfnamefont {F.}~\bibnamefont
  {Khalili-Araghi}}, \bibinfo {author} {\bibfnamefont {E.}~\bibnamefont
  {Tajkhorshid}}, \ and\ \bibinfo {author} {\bibfnamefont {K.}~\bibnamefont
  {Schulten}},\ }\href@noop {} {\bibfield  {journal} {\bibinfo  {journal} {J.
  Chem. Phys.}\ }\textbf {\bibinfo {volume} {119}},\ \bibinfo {pages} {3559}
  (\bibinfo {year} {2003})}\BibitemShut {NoStop}%
\bibitem [{\citenamefont {Minh}\ and\ \citenamefont
  {Chodera}(2011)}]{Minh2011b}%
  \BibitemOpen
  \bibfield  {author} {\bibinfo {author} {\bibfnamefont {D.~D.~L.}\
  \bibnamefont {Minh}}\ and\ \bibinfo {author} {\bibfnamefont {J.~D.}\
  \bibnamefont {Chodera}},\ }\href@noop {} {\bibfield  {journal} {\bibinfo
  {journal} {J. Chem. Phys.}\ }\textbf {\bibinfo {volume} {134}},\ \bibinfo
  {pages} {024111} (\bibinfo {year} {2011})}\BibitemShut {NoStop}%
\bibitem [{\citenamefont {Hummer}(2001)}]{Hummer2001c}%
  \BibitemOpen
  \bibfield  {author} {\bibinfo {author} {\bibfnamefont {G.}~\bibnamefont
  {Hummer}},\ }\href@noop {} {\bibfield  {journal} {\bibinfo  {journal} {J.
  Chem. Phys.}\ }\textbf {\bibinfo {volume} {114}},\ \bibinfo {pages} {7330}
  (\bibinfo {year} {2001})}\BibitemShut {NoStop}%
\bibitem [{\citenamefont {Ytreberg}\ and\ \citenamefont
  {Zuckerman}(2004)}]{Ytreberg2004c}%
  \BibitemOpen
  \bibfield  {author} {\bibinfo {author} {\bibfnamefont {F.~M.}\ \bibnamefont
  {Ytreberg}}\ and\ \bibinfo {author} {\bibfnamefont {D.~M.}\ \bibnamefont
  {Zuckerman}},\ }\href@noop {} {\bibfield  {journal} {\bibinfo  {journal} {J.
  Comput. Chem.}\ }\textbf {\bibinfo {volume} {25}},\ \bibinfo {pages} {1749}
  (\bibinfo {year} {2004})}\BibitemShut {NoStop}%
\bibitem [{\citenamefont {Dellago}\ and\ \citenamefont
  {Hummer}(2014)}]{Dellago2014}%
  \BibitemOpen
  \bibfield  {author} {\bibinfo {author} {\bibfnamefont {C.}~\bibnamefont
  {Dellago}}\ and\ \bibinfo {author} {\bibfnamefont {G.}~\bibnamefont
  {Hummer}},\ }\href@noop {} {\bibfield  {journal} {\bibinfo  {journal}
  {entropy}\ }\textbf {\bibinfo {volume} {16}},\ \bibinfo {pages} {41}
  (\bibinfo {year} {2014})}\BibitemShut {NoStop}%
\bibitem [{\citenamefont {Sandberg}\ \emph {et~al.}(2015)\citenamefont
  {Sandberg}, \citenamefont {Banchelli}, \citenamefont {Guardiani},
  \citenamefont {Menichetti}, \citenamefont {Caminati},\ and\ \citenamefont
  {Procacci}}]{Sandberg2015}%
  \BibitemOpen
  \bibfield  {author} {\bibinfo {author} {\bibfnamefont {R.~B.}\ \bibnamefont
  {Sandberg}}, \bibinfo {author} {\bibfnamefont {M.}~\bibnamefont {Banchelli}},
  \bibinfo {author} {\bibfnamefont {C.}~\bibnamefont {Guardiani}}, \bibinfo
  {author} {\bibfnamefont {S.}~\bibnamefont {Menichetti}}, \bibinfo {author}
  {\bibfnamefont {G.}~\bibnamefont {Caminati}}, \ and\ \bibinfo {author}
  {\bibfnamefont {P.}~\bibnamefont {Procacci}},\ }\href@noop {} {\bibfield
  {journal} {\bibinfo  {journal} {J. Chem. Theory Comput.}\ }\textbf {\bibinfo
  {volume} {11}},\ \bibinfo {pages} {423} (\bibinfo {year} {2015})}\BibitemShut
  {NoStop}%
\bibitem [{\citenamefont {Giovannelli}\ \emph {et~al.}(2017)\citenamefont
  {Giovannelli}, \citenamefont {Procacci}, \citenamefont {Cardini},
  \citenamefont {Pagliai}, \citenamefont {Volkov},\ and\ \citenamefont
  {Chelli}}]{Giovannelli2017}%
  \BibitemOpen
  \bibfield  {author} {\bibinfo {author} {\bibfnamefont {E.}~\bibnamefont
  {Giovannelli}}, \bibinfo {author} {\bibfnamefont {P.}~\bibnamefont
  {Procacci}}, \bibinfo {author} {\bibfnamefont {G.}~\bibnamefont {Cardini}},
  \bibinfo {author} {\bibfnamefont {M.}~\bibnamefont {Pagliai}}, \bibinfo
  {author} {\bibfnamefont {V.}~\bibnamefont {Volkov}}, \ and\ \bibinfo {author}
  {\bibfnamefont {R.}~\bibnamefont {Chelli}},\ }\href@noop {} {\bibfield
  {journal} {\bibinfo  {journal} {J. Chem. Theory Comput.}\ }\textbf {\bibinfo
  {volume} {13}},\ \bibinfo {pages} {5874} (\bibinfo {year}
  {2017})}\BibitemShut {NoStop}%
\bibitem [{\citenamefont {Gore}, \citenamefont {Ritort},\ and\ \citenamefont
  {Bustamante}(2003)}]{Gore2003}%
  \BibitemOpen
  \bibfield  {author} {\bibinfo {author} {\bibfnamefont {J.}~\bibnamefont
  {Gore}}, \bibinfo {author} {\bibfnamefont {F.}~\bibnamefont {Ritort}}, \ and\
  \bibinfo {author} {\bibfnamefont {C.}~\bibnamefont {Bustamante}},\
  }\href@noop {} {\bibfield  {journal} {\bibinfo  {journal} {Proc. Natl. Acad.
  Sci. USA}\ }\textbf {\bibinfo {volume} {100}},\ \bibinfo {pages} {12564}
  (\bibinfo {year} {2003})}\BibitemShut {NoStop}%
\bibitem [{\citenamefont {Jarzynski}(2006)}]{Jarzynski2006}%
  \BibitemOpen
  \bibfield  {author} {\bibinfo {author} {\bibfnamefont {C.}~\bibnamefont
  {Jarzynski}},\ }\href@noop {} {\bibfield  {journal} {\bibinfo  {journal}
  {Phys. Rev. E}\ }\textbf {\bibinfo {volume} {73}},\ \bibinfo {pages} {46105}
  (\bibinfo {year} {2006})}\BibitemShut {NoStop}%
\bibitem [{\citenamefont {Kosztin}, \citenamefont {Barz},\ and\ \citenamefont
  {Janosi}(2006)}]{Kosztin2006a}%
  \BibitemOpen
  \bibfield  {author} {\bibinfo {author} {\bibfnamefont {I.}~\bibnamefont
  {Kosztin}}, \bibinfo {author} {\bibfnamefont {B.}~\bibnamefont {Barz}}, \
  and\ \bibinfo {author} {\bibfnamefont {L.}~\bibnamefont {Janosi}},\
  }\href@noop {} {\bibfield  {journal} {\bibinfo  {journal} {J. Chem. Phys.}\
  }\textbf {\bibinfo {volume} {124}},\ \bibinfo {pages} {064106} (\bibinfo
  {year} {2006})}\BibitemShut {NoStop}%
\bibitem [{\citenamefont {Chelli}\ and\ \citenamefont
  {Procacci}(2009)}]{Chelli2009a}%
  \BibitemOpen
  \bibfield  {author} {\bibinfo {author} {\bibfnamefont {R.}~\bibnamefont
  {Chelli}}\ and\ \bibinfo {author} {\bibfnamefont {P.}~\bibnamefont
  {Procacci}},\ }\href@noop {} {\bibfield  {journal} {\bibinfo  {journal}
  {Phys. Chem. Chem. Phys.}\ }\textbf {\bibinfo {volume} {11}},\ \bibinfo
  {pages} {1152} (\bibinfo {year} {2009})}\BibitemShut {NoStop}%
\bibitem [{\citenamefont {Frey}\ \emph {et~al.}(2015)\citenamefont {Frey},
  \citenamefont {Li}, \citenamefont {Wijeratne},\ and\ \citenamefont
  {Kiang}}]{Frey2015}%
  \BibitemOpen
  \bibfield  {author} {\bibinfo {author} {\bibfnamefont {E.~W.}\ \bibnamefont
  {Frey}}, \bibinfo {author} {\bibfnamefont {J.}~\bibnamefont {Li}}, \bibinfo
  {author} {\bibfnamefont {S.~S.}\ \bibnamefont {Wijeratne}}, \ and\ \bibinfo
  {author} {\bibfnamefont {C.~H.}\ \bibnamefont {Kiang}},\ }\href@noop {}
  {\bibfield  {journal} {\bibinfo  {journal} {J. Phys. Chem. B}\ }\textbf
  {\bibinfo {volume} {119}},\ \bibinfo {pages} {5132} (\bibinfo {year}
  {2015})}\BibitemShut {NoStop}%
\bibitem [{\citenamefont {Ngo}\ \emph {et~al.}(2016)\citenamefont {Ngo},
  \citenamefont {Kim}, \citenamefont {Allen},\ and\ \citenamefont
  {Noskov}}]{Ngo2016}%
  \BibitemOpen
  \bibfield  {author} {\bibinfo {author} {\bibfnamefont {V.~A.}\ \bibnamefont
  {Ngo}}, \bibinfo {author} {\bibfnamefont {I.}~\bibnamefont {Kim}}, \bibinfo
  {author} {\bibfnamefont {T.~W.}\ \bibnamefont {Allen}}, \ and\ \bibinfo
  {author} {\bibfnamefont {S.~Y.}\ \bibnamefont {Noskov}},\ }\href@noop {}
  {\bibfield  {journal} {\bibinfo  {journal} {J. Chem. Theory Comput.}\
  }\textbf {\bibinfo {volume} {12}},\ \bibinfo {pages} {1000} (\bibinfo {year}
  {2016})}\BibitemShut {NoStop}%
\bibitem [{\citenamefont {Calderon}, \citenamefont {Janosi},\ and\
  \citenamefont {Kosztin}(2009)}]{Calderon2009}%
  \BibitemOpen
  \bibfield  {author} {\bibinfo {author} {\bibfnamefont {C.~P.}\ \bibnamefont
  {Calderon}}, \bibinfo {author} {\bibfnamefont {L.}~\bibnamefont {Janosi}}, \
  and\ \bibinfo {author} {\bibfnamefont {I.}~\bibnamefont {Kosztin}},\
  }\href@noop {} {\bibfield  {journal} {\bibinfo  {journal} {J. Chem. Phys.}\
  }\textbf {\bibinfo {volume} {130}},\ \bibinfo {pages} {144908} (\bibinfo
  {year} {2009})}\BibitemShut {NoStop}%
\bibitem [{\citenamefont {Giorgino}\ and\ \citenamefont {{De
  Fabritiis}}(2011)}]{Giorgino2011}%
  \BibitemOpen
  \bibfield  {author} {\bibinfo {author} {\bibfnamefont {T.}~\bibnamefont
  {Giorgino}}\ and\ \bibinfo {author} {\bibfnamefont {G.}~\bibnamefont {{De
  Fabritiis}}},\ }\href@noop {} {\bibfield  {journal} {\bibinfo  {journal} {J.
  Chem. Theory Comput.}\ }\textbf {\bibinfo {volume} {7}},\ \bibinfo {pages}
  {1943} (\bibinfo {year} {2011})}\BibitemShut {NoStop}%
\bibitem [{\citenamefont {Sinha}, \citenamefont {Ganguly},\ and\ \citenamefont
  {Bandyopadhyay}(2012)}]{Sinha2012}%
  \BibitemOpen
  \bibfield  {author} {\bibinfo {author} {\bibfnamefont {V.}~\bibnamefont
  {Sinha}}, \bibinfo {author} {\bibfnamefont {B.}~\bibnamefont {Ganguly}}, \
  and\ \bibinfo {author} {\bibfnamefont {T.}~\bibnamefont {Bandyopadhyay}},\
  }\href@noop {} {\bibfield  {journal} {\bibinfo  {journal} {PLoS ONE}\
  }\textbf {\bibinfo {volume} {7}},\ \bibinfo {pages} {e40188} (\bibinfo {year}
  {2012})}\BibitemShut {NoStop}%
\bibitem [{\citenamefont {Soto-Delgado}, \citenamefont {Tapia},\ and\
  \citenamefont {Torras}(2016)}]{Soto-Delgado2016}%
  \BibitemOpen
  \bibfield  {author} {\bibinfo {author} {\bibfnamefont {J.}~\bibnamefont
  {Soto-Delgado}}, \bibinfo {author} {\bibfnamefont {R.~A.}\ \bibnamefont
  {Tapia}}, \ and\ \bibinfo {author} {\bibfnamefont {J.}~\bibnamefont
  {Torras}},\ }\href@noop {} {\bibfield  {journal} {\bibinfo  {journal} {J.
  Chem. Theory Comput.}\ }\textbf {\bibinfo {volume} {12}},\ \bibinfo {pages}
  {4735} (\bibinfo {year} {2016})}\BibitemShut {NoStop}%
\bibitem [{\citenamefont {Crooks}(1998)}]{Crooks1998}%
  \BibitemOpen
  \bibfield  {author} {\bibinfo {author} {\bibfnamefont {G.~E.}\ \bibnamefont
  {Crooks}},\ }\href@noop {} {\bibfield  {journal} {\bibinfo  {journal} {J.
  Stat. Phys.}\ }\textbf {\bibinfo {volume} {90}},\ \bibinfo {pages} {1481}
  (\bibinfo {year} {1998})}\BibitemShut {NoStop}%
\bibitem [{\citenamefont {Mackerell~Jr.}, \citenamefont {Feig},\ and\
  \citenamefont {Brooks~III}(2004)}]{Mackerell2004}%
  \BibitemOpen
  \bibfield  {author} {\bibinfo {author} {\bibfnamefont {A.~D.}\ \bibnamefont
  {Mackerell~Jr.}}, \bibinfo {author} {\bibfnamefont {M.}~\bibnamefont {Feig}},
  \ and\ \bibinfo {author} {\bibfnamefont {C.~L.}\ \bibnamefont {Brooks~III}},\
  }\href@noop {} {\bibfield  {journal} {\bibinfo  {journal} {J. Comput. Chem.}\
  }\textbf {\bibinfo {volume} {25}},\ \bibinfo {pages} {1400} (\bibinfo {year}
  {2004})}\BibitemShut {NoStop}%
\bibitem [{\citenamefont {MacKerell}\ \emph {et~al.}(1998)\citenamefont
  {MacKerell}, \citenamefont {Bashford}, \citenamefont {Bellott}, \citenamefont
  {Dunbrack}, \citenamefont {Evanseck}, \citenamefont {Field}, \citenamefont
  {Fischer}, \citenamefont {Gao}, \citenamefont {Guo}, \citenamefont {Ha},
  \citenamefont {Joseph-McCarthy}, \citenamefont {Kuchnir}, \citenamefont
  {Kuczera}, \citenamefont {Lau}, \citenamefont {Mattos}, \citenamefont
  {Michnick}, \citenamefont {Ngo}, \citenamefont {Nguyen}, \citenamefont
  {Prodhom}, \citenamefont {Reiher}, \citenamefont {Roux}, \citenamefont
  {Schlenkrich}, \citenamefont {Smith}, \citenamefont {Stote}, \citenamefont
  {Straub}, \citenamefont {Watanabe}, \citenamefont {Wiórkiewicz-Kuczera},
  \citenamefont {Yin},\ and\ \citenamefont {Karplus}}]{MacKerell1998}%
  \BibitemOpen
  \bibfield  {author} {\bibinfo {author} {\bibfnamefont {A.~D.}\ \bibnamefont
  {MacKerell}}, \bibinfo {author} {\bibfnamefont {D.}~\bibnamefont {Bashford}},
  \bibinfo {author} {\bibfnamefont {M.}~\bibnamefont {Bellott}}, \bibinfo
  {author} {\bibfnamefont {R.~L.}\ \bibnamefont {Dunbrack}}, \bibinfo {author}
  {\bibfnamefont {J.~D.}\ \bibnamefont {Evanseck}}, \bibinfo {author}
  {\bibfnamefont {M.~J.}\ \bibnamefont {Field}}, \bibinfo {author}
  {\bibfnamefont {S.}~\bibnamefont {Fischer}}, \bibinfo {author} {\bibfnamefont
  {J.}~\bibnamefont {Gao}}, \bibinfo {author} {\bibfnamefont {H.}~\bibnamefont
  {Guo}}, \bibinfo {author} {\bibfnamefont {S.}~\bibnamefont {Ha}}, \bibinfo
  {author} {\bibfnamefont {D.}~\bibnamefont {Joseph-McCarthy}}, \bibinfo
  {author} {\bibfnamefont {L.}~\bibnamefont {Kuchnir}}, \bibinfo {author}
  {\bibfnamefont {K.}~\bibnamefont {Kuczera}}, \bibinfo {author} {\bibfnamefont
  {F.~T.~K.}\ \bibnamefont {Lau}}, \bibinfo {author} {\bibfnamefont
  {C.}~\bibnamefont {Mattos}}, \bibinfo {author} {\bibfnamefont
  {S.}~\bibnamefont {Michnick}}, \bibinfo {author} {\bibfnamefont
  {T.}~\bibnamefont {Ngo}}, \bibinfo {author} {\bibfnamefont {D.~T.}\
  \bibnamefont {Nguyen}}, \bibinfo {author} {\bibfnamefont {B.}~\bibnamefont
  {Prodhom}}, \bibinfo {author} {\bibfnamefont {W.~E.}\ \bibnamefont {Reiher}},
  \bibinfo {author} {\bibfnamefont {B.}~\bibnamefont {Roux}}, \bibinfo {author}
  {\bibfnamefont {M.}~\bibnamefont {Schlenkrich}}, \bibinfo {author}
  {\bibfnamefont {J.~C.}\ \bibnamefont {Smith}}, \bibinfo {author}
  {\bibfnamefont {R.}~\bibnamefont {Stote}}, \bibinfo {author} {\bibfnamefont
  {J.}~\bibnamefont {Straub}}, \bibinfo {author} {\bibfnamefont
  {M.}~\bibnamefont {Watanabe}}, \bibinfo {author} {\bibfnamefont
  {J.}~\bibnamefont {Wiórkiewicz-Kuczera}}, \bibinfo {author} {\bibfnamefont
  {D.}~\bibnamefont {Yin}}, \ and\ \bibinfo {author} {\bibfnamefont
  {M.}~\bibnamefont {Karplus}},\ }\href@noop {} {\bibfield  {journal} {\bibinfo
   {journal} {J. Phys. Chem. B}\ }\textbf {\bibinfo {volume} {102}},\ \bibinfo
  {pages} {3586} (\bibinfo {year} {1998})}\BibitemShut {NoStop}%
\bibitem [{\citenamefont {MacKerell}, \citenamefont {Feig},\ and\ \citenamefont
  {Brooks}(2004)}]{MacKerell2004b}%
  \BibitemOpen
  \bibfield  {author} {\bibinfo {author} {\bibfnamefont {A.~D.}\ \bibnamefont
  {MacKerell}}, \bibinfo {author} {\bibfnamefont {M.}~\bibnamefont {Feig}}, \
  and\ \bibinfo {author} {\bibfnamefont {C.~L.}\ \bibnamefont {Brooks}},\
  }\href@noop {} {\bibfield  {journal} {\bibinfo  {journal} {J Am. Chem. Soc.}\
  }\textbf {\bibinfo {volume} {126}},\ \bibinfo {pages} {698} (\bibinfo {year}
  {2004})}\BibitemShut {NoStop}%
\bibitem [{\citenamefont {Phillips}\ \emph {et~al.}(2005)\citenamefont
  {Phillips}, \citenamefont {Braun}, \citenamefont {Wang}, \citenamefont
  {Gumbart}, \citenamefont {Tajkhorshid}, \citenamefont {Villa}, \citenamefont
  {Chipot}, \citenamefont {Skeel}, \citenamefont {Kalé},\ and\ \citenamefont
  {Schulten}}]{Phillips2005}%
  \BibitemOpen
  \bibfield  {author} {\bibinfo {author} {\bibfnamefont {J.~C.}\ \bibnamefont
  {Phillips}}, \bibinfo {author} {\bibfnamefont {R.}~\bibnamefont {Braun}},
  \bibinfo {author} {\bibfnamefont {W.}~\bibnamefont {Wang}}, \bibinfo {author}
  {\bibfnamefont {J.}~\bibnamefont {Gumbart}}, \bibinfo {author} {\bibfnamefont
  {E.}~\bibnamefont {Tajkhorshid}}, \bibinfo {author} {\bibfnamefont
  {E.}~\bibnamefont {Villa}}, \bibinfo {author} {\bibfnamefont
  {C.}~\bibnamefont {Chipot}}, \bibinfo {author} {\bibfnamefont {R.~D.}\
  \bibnamefont {Skeel}}, \bibinfo {author} {\bibfnamefont {L.}~\bibnamefont
  {Kalé}}, \ and\ \bibinfo {author} {\bibfnamefont {K.}~\bibnamefont
  {Schulten}},\ }\href@noop {} {\bibfield  {journal} {\bibinfo  {journal} {J.
  Comput. Chem.}\ }\textbf {\bibinfo {volume} {26}},\ \bibinfo {pages} {1781}
  (\bibinfo {year} {2005})}\BibitemShut {NoStop}%
\bibitem [{\citenamefont {Park}, \citenamefont {Khalili},\ and\ \citenamefont
  {Strumpfer}(2012)}]{Park2012}%
  \BibitemOpen
  \bibfield  {author} {\bibinfo {author} {\bibfnamefont {S.}~\bibnamefont
  {Park}}, \bibinfo {author} {\bibfnamefont {F.}~\bibnamefont {Khalili}}, \
  and\ \bibinfo {author} {\bibfnamefont {J.}~\bibnamefont {Strumpfer}},\ }\href
  {http://www.ks.uiuc.edu/Training/Tutorials/} {\enquote {\bibinfo {title}
  {Stretching deca-alanine, http://www.ks.uiuc.edu/training/tutorials/},}\ }
  (\bibinfo {year} {2012})\BibitemShut {NoStop}%
\bibitem [{\citenamefont {Velez-Vega}\ and\ \citenamefont
  {Gilson}(2013)}]{Velez-Vega2013}%
  \BibitemOpen
  \bibfield  {author} {\bibinfo {author} {\bibfnamefont {C.}~\bibnamefont
  {Velez-Vega}}\ and\ \bibinfo {author} {\bibfnamefont {M.~K.}\ \bibnamefont
  {Gilson}},\ }\href@noop {} {\bibfield  {journal} {\bibinfo  {journal} {J.
  Comput. Chem.}\ }\textbf {\bibinfo {volume} {34}},\ \bibinfo {pages} {2360}
  (\bibinfo {year} {2013})}\BibitemShut {NoStop}%
\bibitem [{\citenamefont {Nguyen}\ and\ \citenamefont
  {Minh}(2016)}]{Nguyen2016}%
  \BibitemOpen
  \bibfield  {author} {\bibinfo {author} {\bibfnamefont {T.~H.}\ \bibnamefont
  {Nguyen}}\ and\ \bibinfo {author} {\bibfnamefont {D.~D.~L.}\ \bibnamefont
  {Minh}},\ }\href@noop {} {\bibfield  {journal} {\bibinfo  {journal} {J. Chem.
  Theory Comput.}\ }\textbf {\bibinfo {volume} {12}},\ \bibinfo {pages} {2154}
  (\bibinfo {year} {2016})}\BibitemShut {NoStop}%
\bibitem [{\citenamefont {Gilson}, \citenamefont {Gilson},\ and\ \citenamefont
  {Potter}(2003)}]{Gilson2003}%
  \BibitemOpen
  \bibfield  {author} {\bibinfo {author} {\bibfnamefont {M.~K.}\ \bibnamefont
  {Gilson}}, \bibinfo {author} {\bibfnamefont {H.~S.~R.}\ \bibnamefont
  {Gilson}}, \ and\ \bibinfo {author} {\bibfnamefont {M.~J.}\ \bibnamefont
  {Potter}},\ }\href@noop {} {\bibfield  {journal} {\bibinfo  {journal} {J.
  Chem. Inf. Comput. Sci.}\ }\textbf {\bibinfo {volume} {43}},\ \bibinfo
  {pages} {1982} (\bibinfo {year} {2003})}\BibitemShut {NoStop}%
\bibitem [{\citenamefont {Hornak}\ \emph {et~al.}(2006)\citenamefont {Hornak},
  \citenamefont {Abel}, \citenamefont {Okur}, \citenamefont {Strockbine},
  \citenamefont {Roitberg},\ and\ \citenamefont {Simmerling}}]{Hornak2006}%
  \BibitemOpen
  \bibfield  {author} {\bibinfo {author} {\bibfnamefont {V.}~\bibnamefont
  {Hornak}}, \bibinfo {author} {\bibfnamefont {R.}~\bibnamefont {Abel}},
  \bibinfo {author} {\bibfnamefont {A.}~\bibnamefont {Okur}}, \bibinfo {author}
  {\bibfnamefont {B.}~\bibnamefont {Strockbine}}, \bibinfo {author}
  {\bibfnamefont {A.}~\bibnamefont {Roitberg}}, \ and\ \bibinfo {author}
  {\bibfnamefont {C.}~\bibnamefont {Simmerling}},\ }\href@noop {} {\bibfield
  {journal} {\bibinfo  {journal} {Proteins: Struct., Funct., Bioinf.}\ }\textbf
  {\bibinfo {volume} {65}},\ \bibinfo {pages} {712} (\bibinfo {year}
  {2006})}\BibitemShut {NoStop}%
\bibitem [{\citenamefont {Wang}\ \emph {et~al.}(2004)\citenamefont {Wang},
  \citenamefont {Wolf}, \citenamefont {Caldwell}, \citenamefont {Kollman},\
  and\ \citenamefont {Case}}]{Wang2004a}%
  \BibitemOpen
  \bibfield  {author} {\bibinfo {author} {\bibfnamefont {J.}~\bibnamefont
  {Wang}}, \bibinfo {author} {\bibfnamefont {R.~M.}\ \bibnamefont {Wolf}},
  \bibinfo {author} {\bibfnamefont {J.~W.}\ \bibnamefont {Caldwell}}, \bibinfo
  {author} {\bibfnamefont {P.~A.}\ \bibnamefont {Kollman}}, \ and\ \bibinfo
  {author} {\bibfnamefont {D.~A.}\ \bibnamefont {Case}},\ }\href@noop {}
  {\bibfield  {journal} {\bibinfo  {journal} {J. Comput. Chem.}\ }\textbf
  {\bibinfo {volume} {25}},\ \bibinfo {pages} {1157} (\bibinfo {year}
  {2004})}\BibitemShut {NoStop}%
\bibitem [{\citenamefont {Vainio}\ and\ \citenamefont
  {Johnson}(2007)}]{Vainio2007}%
  \BibitemOpen
  \bibfield  {author} {\bibinfo {author} {\bibfnamefont {M.~J.}\ \bibnamefont
  {Vainio}}\ and\ \bibinfo {author} {\bibfnamefont {M.~S.}\ \bibnamefont
  {Johnson}},\ }\href@noop {} {\bibfield  {journal} {\bibinfo  {journal} {J.
  Chem. Inf. Model.}\ }\textbf {\bibinfo {volume} {47}},\ \bibinfo {pages}
  {2462} (\bibinfo {year} {2007})}\BibitemShut {NoStop}%
\bibitem [{\citenamefont {Wang}\ \emph {et~al.}(2006)\citenamefont {Wang},
  \citenamefont {Wang}, \citenamefont {Kollman},\ and\ \citenamefont
  {Case}}]{Wang2006}%
  \BibitemOpen
  \bibfield  {author} {\bibinfo {author} {\bibfnamefont {J.}~\bibnamefont
  {Wang}}, \bibinfo {author} {\bibfnamefont {W.}~\bibnamefont {Wang}}, \bibinfo
  {author} {\bibfnamefont {P.~A.}\ \bibnamefont {Kollman}}, \ and\ \bibinfo
  {author} {\bibfnamefont {D.~A.}\ \bibnamefont {Case}},\ }\href@noop {}
  {\bibfield  {journal} {\bibinfo  {journal} {J. Mol. Graph. Model.}\ }\textbf
  {\bibinfo {volume} {25}},\ \bibinfo {pages} {247} (\bibinfo {year}
  {2006})}\BibitemShut {NoStop}%
\bibitem [{\citenamefont {Smith}\ and\ \citenamefont {Dang}(1994)}]{Smith1994}%
  \BibitemOpen
  \bibfield  {author} {\bibinfo {author} {\bibfnamefont {D.~E.}\ \bibnamefont
  {Smith}}\ and\ \bibinfo {author} {\bibfnamefont {L.~X.}\ \bibnamefont
  {Dang}},\ }\href@noop {} {\bibfield  {journal} {\bibinfo  {journal} {J. Chem.
  Phys.}\ }\textbf {\bibinfo {volume} {100}},\ \bibinfo {pages} {3757}
  (\bibinfo {year} {1994})}\BibitemShut {NoStop}%
\bibitem [{\citenamefont {Shirts}\ and\ \citenamefont
  {Chodera}(2008)}]{Shirts2008}%
  \BibitemOpen
  \bibfield  {author} {\bibinfo {author} {\bibfnamefont {M.~R.}\ \bibnamefont
  {Shirts}}\ and\ \bibinfo {author} {\bibfnamefont {J.~D.}\ \bibnamefont
  {Chodera}},\ }\href@noop {} {\bibfield  {journal} {\bibinfo  {journal} {J.
  Chem. Phys.}\ }\textbf {\bibinfo {volume} {129}},\ \bibinfo {pages} {124105}
  (\bibinfo {year} {2008})}\BibitemShut {NoStop}%
\bibitem [{\citenamefont {Ferrenberg}\ and\ \citenamefont
  {Swendsen}(1989)}]{Ferrenberg1989}%
  \BibitemOpen
  \bibfield  {author} {\bibinfo {author} {\bibfnamefont {A.~M.}\ \bibnamefont
  {Ferrenberg}}\ and\ \bibinfo {author} {\bibfnamefont {R.~H.}\ \bibnamefont
  {Swendsen}},\ }\href@noop {} {\bibfield  {journal} {\bibinfo  {journal}
  {Phys. Rev. Lett.}\ }\textbf {\bibinfo {volume} {63}},\ \bibinfo {pages}
  {1195} (\bibinfo {year} {1989})}\BibitemShut {NoStop}%
\bibitem [{\citenamefont {Kumar}\ \emph {et~al.}(1992)\citenamefont {Kumar},
  \citenamefont {Rosenberg}, \citenamefont {Bouzida}, \citenamefont
  {Swendsen},\ and\ \citenamefont {Kollman}}]{Kumar1992}%
  \BibitemOpen
  \bibfield  {author} {\bibinfo {author} {\bibfnamefont {S.}~\bibnamefont
  {Kumar}}, \bibinfo {author} {\bibfnamefont {J.~M.}\ \bibnamefont
  {Rosenberg}}, \bibinfo {author} {\bibfnamefont {D.}~\bibnamefont {Bouzida}},
  \bibinfo {author} {\bibfnamefont {R.~H.}\ \bibnamefont {Swendsen}}, \ and\
  \bibinfo {author} {\bibfnamefont {P.~A.}\ \bibnamefont {Kollman}},\
  }\href@noop {} {\bibfield  {journal} {\bibinfo  {journal} {J. Comput. Chem.}\
  }\textbf {\bibinfo {volume} {13}},\ \bibinfo {pages} {1011} (\bibinfo {year}
  {1992})}\BibitemShut {NoStop}%
\end{thebibliography}

%

\end{document}